\documentclass[lettersize,journal]{IEEEtran}

\ifCLASSINFOpdf
\else
\fi
\usepackage{pgfplots}
\pgfplotsset{compat=1.18}
\usepackage{cite}
\usepackage{amsmath,amssymb,amsfonts,amsthm}

\usepackage{algorithm}
\usepackage{array}
\usepackage{algpseudocode}
\usepackage{graphicx}
\usepackage{textcomp}
\usepackage{booktabs}
\usepackage{tikz}
\usepackage{textcomp}
\usepackage{stfloats}
\usepackage{url}
\usepackage{verbatim}
\usetikzlibrary{arrows, shapes, positioning}
\usepackage{bm}
\usepackage{xcolor}
\usepackage{color}

\def\BibTeX{{\rm B\kern-.05em{\sc i\kern-.025em b}\kern-.08em
    T\kern-.1667em\lower.7ex\hbox{E}\kern-.125emX}}
\usepackage{verbatim}
\usepackage{kotex}
\usepackage{dsfont}
\usepackage{subcaption}
\usepackage{mathtools}
\usepackage{multirow}
\usepackage{makecell}

\newcommand\figref{Figure~\ref}
\newtheorem{theorem}{Theorem}
\newtheorem{remark}{Remark}

\newtheorem{proposition}{Proposition}

\usepackage{makecell}
\usepackage{dcolumn}
\newcommand{\tblue}[1]{\textcolor{black}{#1}}
\newcolumntype{d}[1]{D{.}{.}{#1}}

\makeatletter
\newcommand{\ostar}{\mathbin{\mathpalette\make@circled\star}}
\newcommand{\make@circled}[2]{%
  \ooalign{$\m@th#1\smallbigcirc{#1}$\cr\hidewidth$\m@th#1#2$\hidewidth\cr}%
}
\newcommand{\smallbigcirc}[1]{%
  \vcenter{\hbox{\scalebox{0.77778}{$\m@th#1\bigcirc$}}}%
}
\makeatother

\begin{document}


\title{\fontsize{19}{24}\selectfont Communication-Efficient Hybrid Language Model via Uncertainty-Aware Opportunistic and Compressed Transmission}


\author{Seungeun~Oh, Jinhyuk~Kim, Jihong~Park, Seung-Woo~Ko, Jinho Choi, Tony Q. S. Quek, and~Seong-Lyun~Kim 
\thanks{S. Oh was with the School of Electrical and Electronic Engineering, Yonsei University, South Korea. He is now with the Information Systems Technology and Design pillar, Singapore University of Technology and Design, Singapore 487372 (e-mail: seungeun\_oh@sutd.edu.sg).}
\thanks{J. Kim and S.-L. Kim are with the School of Electrical and Electronic Engineering, Yonsei University, South Korea (e-mail: \{jh.kim, slkim\}@ramo.yonsei.ac.kr).}
\thanks{J. Park is with the Information Systems Technology and Design pillar, Singapore University of Technology and Design, Singapore 487372 (e-mail: jihong\_park@sutd.edu.sg).}
\thanks{S.-W. Ko is with the Department of Smart Mobility Engineering, Inha University, South Korea (e-mail: swko@inha.ac.kr).}
\thanks{J. Choi is with the School of Electrical and Mechanical Engineering, The University of Adelaide, Australia (e-mail: jinho.choi@adelaide.edu.au).}
\thanks{T. Q. S. Quek is with the Singapore University of Technology and Design, Singapore 487372 (e-mail: tonyquek@sutd.edu.sg).}
}

\markboth{Journal of \LaTeX\ Class Files,~Vol.~X, No.~Y, March~2026}%
{Oh \MakeLowercase{\textit{et al.}}: Communication-Efficient Hybrid Language Model via Uncertainty-Aware Opportunistic and Compressed Transmission}

\maketitle
\vspace{-20pt}

\begin{abstract}
\noindent
To support emerging language-based applications using dispersed and heterogeneous computing resources, the hybrid language model (HLM) offers a promising architecture, where an on-device small language model (SLM) generates draft tokens that are validated and corrected by a remote large language model (LLM). However, the original HLM suffers from substantial communication overhead, as the LLM requires the SLM to upload the full vocabulary distribution for each token. Moreover, both communication and computation resources are wasted when the LLM validates tokens that are highly likely to be accepted. To overcome these limitations, we propose \textit{communication-efficient and uncertainty-aware HLM (CU-HLM)}. In CU-HLM, the SLM transmits truncated vocabulary distributions only when its output uncertainty is high. We validate the feasibility of this opportunistic transmission by discovering a strong correlation between SLM's uncertainty and LLM's rejection probability. Furthermore, we theoretically derive optimal uncertainty thresholds and optimal vocabulary truncation strategies. Simulation results show that, compared to standard HLM, CU-HLM achieves up to \textbf{206}$\times$ higher token throughput by skipping \textbf{74.8\%} transmissions with \textbf{97.4\%} vocabulary compression, while maintaining \textbf{97.4\%} accuracy.
\end{abstract}



\vspace{-10pt}

\IEEEpeerreviewmaketitle

\vspace{-10pt}
\section{Introduction}

\IEEEPARstart{L}{arge} language models (LLMs), with their massive parameter counts and rich training data, have demonstrated remarkable emergent capabilities~\cite{hoffmann2022training}. These capabilities span a wide range of applications, including open-domain question answering, code generation, commonsense reasoning, and even robotic control~\cite{thirunavukarasu2023large,demszky2023using,yang2024harnessing}. To seamlessly adopt LLMs into a wireless edge environment, the hybrid language model (HLM) framework~\cite{hao2024hybrid} has emerged, which physically splits the inference task between an on-device small language model (SLM) and a remote LLM. Specifically, given a token—a basic unit of text generation such as a word or subword fragment—and the corresponding vocabulary, which denotes the full set of possible tokens the model can output, the SLM proposes a draft token based on a given prompt, which is then transmitted to the server. The server-side LLM then decides whether to accept or resample it. This speculative decoding approach~\cite{leviathan2023fast} ensures that the output distribution matches that of the LLM, preserving inference accuracy while relieving the LLM's computation burden.

Despite these advantages, HLM suffers from severely limited token throughput due to its architectural overhead. According to our simulations, to be detailed in Section~\ref{sec:eval}, each token generation requires:
(i) SLM-to-LLM uplink transmission of the full 32{,}000-token vocabulary distribution, incurring up to 92\,kB of payload per token; and
(ii) execution of both SLM and LLM inference, which take 25.6\,ms and 104.6\,ms, respectively—limiting the overall token throughput to fewer than 5 tokens per second under 10\,MHz uplink bandwidth.

\begin{figure}[t]
\centering
\includegraphics[width=0.5\textwidth]{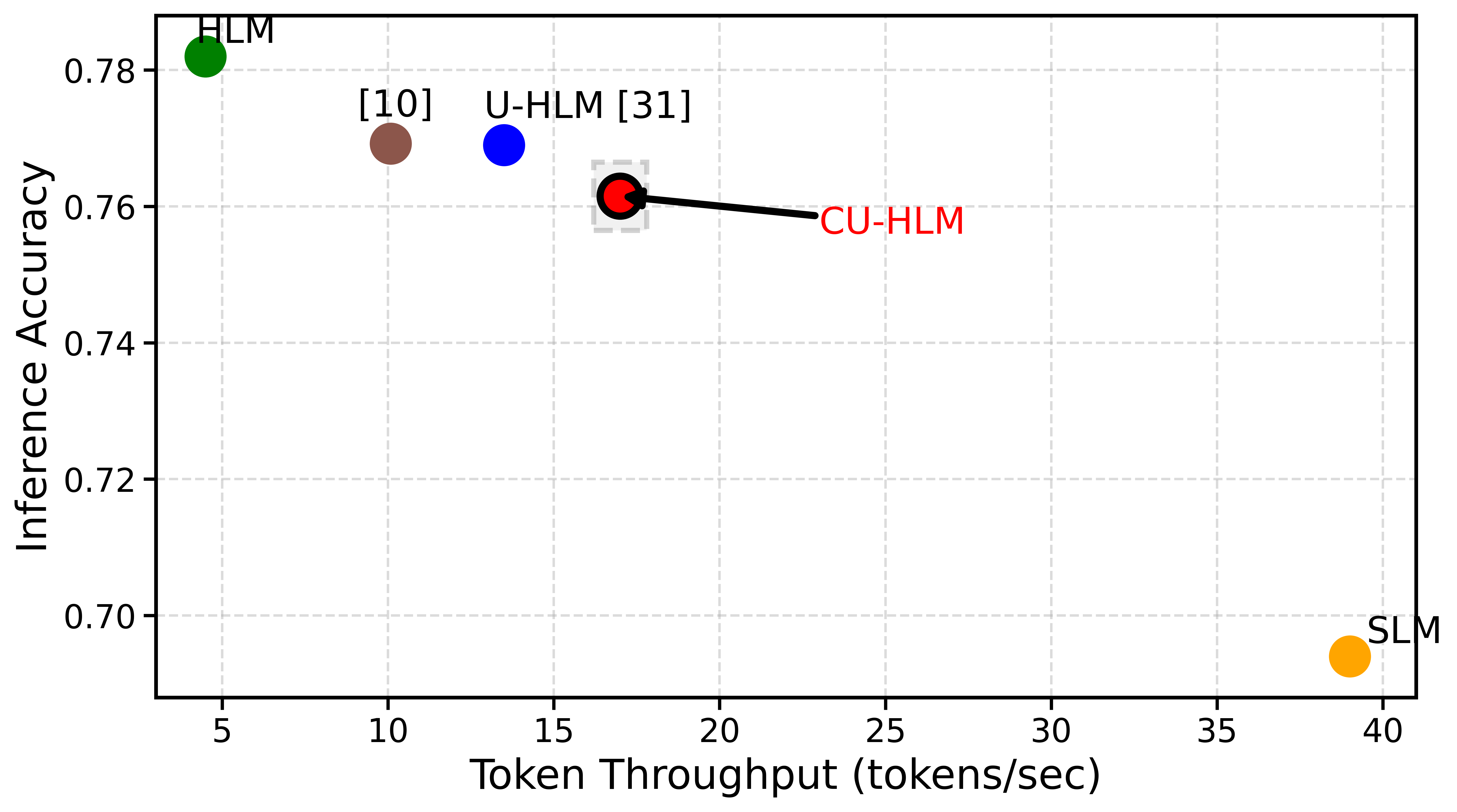}
\caption{\tblue{Comparison of inference accuracy and token throughput, where token throughput is inversely proportional to end-to-end latency, including communication and computation delays (SNR 10\,dB, Rayleigh fading).}}
\label{fig:system_model}
\vspace{-10pt}
\end{figure}

To address the token throughput bottleneck inherent in HLMs, in this paper, we propose communication-efficient and uncertainty-aware HLM (\textbf{CU-HLM}) that reduces uplink communication and server-side computation overhead. CU-HLM introduces two key innovations: \textit{uncertainty-aware opportunistic transmission}, which skips both uplink transmission and remote LLM computation when the SLM’s uncertainty is below a threshold; and \textit{uncertainty-aware compressed vocabulary transmission}, which transmits only the top-$k$ token probabilities, with $k$ increasing adaptively with the SLM's uncertainty. These methods were motivated by our experimental findings: a strong linear correlation between the SLM’s uncertainty and the LLM’s rejection probability; and the optimal $k$ increasing with the token’s uncertainty. Leveraging these uncertainty-aware methods with our theoretically derived optimal $k$ and uncertainty threshold, CU-HLM achieves an effective trade-off between token throughput and inference accuracy, as illustrated in \figref{fig:system_model}. 

\begin{table*}[t]
\centering
\caption{\tblue{Survey of communication-efficient hybrid language model frameworks.}}
\label{tab:sd_hlm_comparison}
\setlength{\tabcolsep}{4pt} 
\renewcommand{\arraystretch}{1.1}
\begin{tabular}{p{5.3cm}cccc}
\hline
\textbf{Method} &
\textbf{Token-Level} &
\textbf{Uncertainty} &
\textbf{Opportunistic Tx} &
\textbf{Compression} \\
\hline
LLM-based acceptance~\cite{hao2024hybrid}
& Y & N & Y & N \\
MLP / RL-based forwarding~\cite{she2025token,zheng2025citer}
& Y & N &  Y & N \\
FL-enabled HLM~\cite{solat2025federated}
& Y & Y &  Y & N \\
Top-$k$ truncation~\cite{zheng2025communication,zheng2026fast}
& Y & N & N & Y \\
Client-side verification~\cite{ning2025dssd}
& Y & N &  N & N \\
FlowSpec~\cite{liu2025flowspec}, Ghidorah~\cite{wei2025ghidorah}, Hetero SD~\cite{zhu2025efficient}
& Y & N &  N & N \\
\textbf{CU-HLM (Ours)}
& Y & Y &  Y & Y \\
\hline
\end{tabular}
\vspace{-10pt}
\end{table*}

\vspace{-10pt}
\subsection{Related Works}
LLMs such as OpenAI's GPT~\cite{achiam2023gpt} and DeepSeek~\cite{guo2025deepseek} have demonstrated strong performance across a variety of tasks, including reasoning, information retrieval, and dialogue generation~\cite{yang2024harnessing}. However, their high computational complexity and memory demands make them impractical for on-device deployment~\cite{zhang2024fast, kandala2024tinyllm}. As a result, LLMs are typically hosted on remote servers, leading to considerable inference latency over wireless networks. This limitation has motivated the development of SLMs, such as TinyLlama~\cite{zhang2024tinyllama} and Phi-2~\cite{javaheripi2023phi}, as well as quantized LLMs~\cite{ray2024llmedge}, which enable lightweight inference that can be executed directly on mobile devices. To further enhance the effectiveness of these models, numerous techniques such as model pruning, quantization, and knowledge distillation have been explored~\cite{ma2023llm, li2023loftq, zhou2023distillspec}. Nevertheless, these approaches often come at the cost of reduced inference accuracy~\cite{wang2024model, xu2023llmcad}, underscoring the trade-off between latency and accuracy in mobile AI deployment.

To address the latency-accuracy trade-off in wireless edge inference, HLMs have been considered an attractive solution by leveraging both device-side and server-side computing resources \cite{hao2024hybrid,she2025token,ray2024llmedge,xu2024edgellm}. In the HLM framework, a SLM is deployed on the device, and a LLM resides on the network edge server. Inspired by speculative inference~\cite{leviathan2023fast, chen2023accelerating}, the HLM operates in two stages: the on-device SLM first generates a draft token in response to the user query, and the server-side LLM subsequently verifies whether the token should be accepted or resampled. This collaborative inference process ensures that the resulting output distribution is consistent with that of the LLM alone, thereby preserving accuracy while distributing the computational load.

However, the HLM architecture causes additional communication and computation overheads. Specifically, for each token, the SLM must transmit its full vocabulary distribution to the server to enable verification or correction. In addition, both the SLM and LLM must perform inference for every token generation. This communication and dual computation requirement brings about considerable latency, limiting token throughput. The conventional approaches focusing on computation aspects, such as quantization, model compression, and optimized CPU/GPU memory allocation~\cite{yu2024edge, huang2024edgellm, shi2024inferflow, ray2024llmedge}, can be ineffective since the issue of communication burden is ignored. To address this, early studies such as~\cite{wang2023tabi} propose query-level offloading to address the issue, where the entire input is directed either to the on-device SLM or to the server-side LLM based on the estimated difficulty of the query. While simple and effective at a coarse level, such approaches lack the granularity needed for token-level control. 

Recent research moves towards joint optimization of communication and computation by selectively determining which tokens should be sent to the LLM for verification. For instance, \cite{hao2024hybrid} computes the acceptance probability of each token using the LLM, but this approach still requires full LLM computation for every token. Other works leverage auxiliary AI techniques—such as multi-layer perceptrons (MLPs) \cite{she2025token} or reinforcement learning \cite{zheng2025citer}—to determine whether individual tokens should be forwarded to the server. 
\tblue{Federated learning--enabled HLMs~\cite{solat2025federated} further extend this line of work by building upon uncertainty-based opportunistic skipping mechanisms, including our previous work~\cite{oh2024uncertainty}, and learning uncertainty thresholds across clients in a privacy-preserving manner to adapt token forwarding decisions.
}
While these methods offer flexible token-level control, they incur additional training computation and require architectural modifications, potentially limiting their feasibility in real-time or resource-constrained edge deployments.

\tblue{Several recent works have explicitly examined communication efficiency within the HLM framework. To reduce uplink communication overhead, prior studies have explored top-$k$ vocabulary truncation strategies~\cite{zheng2025communication,zheng2026fast}. Another approach shifts part of the verification process to the client, trading uplink communication for downlink transmission~\cite{ning2025dssd}. However, these communication-centric approaches typically rely on heuristic or fixed design choices, without a rigorous theoretical characterization of the consequences of such communication–computation trade-offs.}

Building upon these prior works, we aim to improve HLM throughput by jointly optimizing communication and computation, while maintaining inference accuracy.
\tblue{This contribution is supported by an uncertainty-centric theoretical foundation that enables flexible and adaptive token-level communication–computation co-design without additional computational burden, as summarized in Table~\ref{tab:sd_hlm_comparison}.}

\vspace{-10pt}
\subsection{Contributions and Paper Organization}
The main contributions of this paper are outlined below:
\begin{itemize}

    \item We propose \textbf{CU-HLM}, which applies opportunistic and compressed vocabulary transmission based on the SLM's uncertainty. The rationale behind this design is supported by empirically discovering strong correlations among the SLM's uncertainty, the LLM's rejection probability, and the vocabulary compression size.

    

    \item \tblue{For uncertainty-aware opportunistic transmission, we theoretically derive an uncertainty threshold that bounds the per-token rejection risk (false positives) under skipping. For uncertainty-aware vocabulary compression, we derive a theoretical upper bound on the total variation distance (TVD) of the resampling distribution, thereby characterizing the distributional distortion introduced by vocabulary truncation and determining the minimum compression size.}


    \item To apply these optimal uncertainty threshold and vocabulary compression, the SLM requires access to the LLM's outputs. Assuming known LLM output statistics, we propose \textbf{CU-HLM (Offline)}, where the vocabulary compression size is fixed based on an average bias constraint. By relaxing the bias upper bound to remove LLM dependency, we further develop \textbf{CU-HLM (Online)}, which dynamically adjusts the vocabulary size according to each token-level bias constraint. 


    \item Extensive simulations across wireless fading channels, SNRs, datasets, and model configurations show that CU-HLM (Online) achieves \textbf{97.4\%} of HLM accuracy and up to \textbf{206$\times$} higher token throughput under poor wireless channel conditions, while reducing uplink transmissions by \textbf{74.8\%} (85.7\% of them are finally accepted) and transmitting only \textbf{2.6\%} of the full vocabulary payload. 

\end{itemize}


Note that the conference version~\cite{oh2024uncertainty} of this work presents only the uncertainty-aware opportunistic transmission, termed U-HLM. To further improve communication efficiency and token throughput, this work additionally introduces uncertainty-aware vocabulary compression, which poses new challenges in optimizing vocabulary compression size and designing online/offline strategies. By theoretically deriving the optimal settings, the proposed CU-HLM (Online) achieves 17.6\% higher token throughput and a 177.5$\times$ communication latency reduction, compared to U-HLM in \cite{oh2024uncertainty}. 
 
The remainder of this paper is organized as follows. Section~\ref{sec:system} presents a systematic overview of the HLM framework, including its integration with wireless communication systems and the definition of token throughput.  
Section~\ref{sec:U-HLM} explores the relationship between token-level uncertainty and LLM rejection probability, and introduces U-HLM by extending the HLM framework with uncertainty-aware opportunistic transmission along with a theoretically derived optimal uncertainty threshold. Section~\ref{sec:CU-HLM} then proposes CU-HLM by integrating a vocabulary compression scheme with U-HLM, and describes its offline and online variants based on optimal vocabulary size selection. Section~\ref{sec:eval} provides a comprehensive empirical evaluation, examining the effects of uncertainty thresholds and vocabulary sizes, validating CU-HLM across various settings, and including an ablation study on the impact of SLM architecture and alignment on token throughput. Finally, Section~\ref{sec:con} concludes the paper.

\vspace{-10pt}
\section{System Model} \label{sec:system}
The network considered in this study comprises a single device and a base station (BS) equipped with a powerful server. In accordance with the HLM architecture proposed in \cite{hao2024hybrid}, the device and server are respectively equipped with a SLM and a LLM. Both the SLM and LLM operate on basic units called tokens and share a common vocabulary $\mathcal{V}$, which defines the complete set of possible tokens. The task is to generate response tokens, given an input token sequence denoted by $\mathbf{s}$, by sampling from the model’s vocabulary distribution.

\subsection{Token Generation of a Hybrid Language Model} \label{section2_A}
The core inference mechanism in the HLM consists of two stages: (1) draft token generation by the SLM, and (2) verification and potential correction by the LLM through resampling. Further details are provided below.

\noindent \textbf{Step 1: SLM's Draft Generation.} \quad
In the $t$-th round, the input to the SLM, denoted by $\mathbf{s}(t)$, is constructed by concatenating the input token sequence $\mathbf{s}$ with the cumulative response token sequence up to the ($t-1$)-th round, denoted by $\mathbf{r}(t{-}1)$: $\mathbf{s}(t) \coloneqq \mathbf{s} \oplus \mathbf{r}(t{-}1),$
where $\oplus$ denotes the concatenation operator.

The SLM processes $\mathbf{s}(t)$ and generates a logit vector $\mathbf{z}(t) = [z_{1}(t), z_{2}(t), \ldots, z_{|\mathcal{V}|}(t)]^\top$.
This logit vector is then normalized to obtain the SLM's vocabulary distribution $\mathbf{x}(t) = [x_{1}(t),x_{2}(t),...,x_{|\mathcal{V}|}(t)]^\top$, where:
\begin{align}
x_v(t) = \frac{\exp(z_v(t))}{\sum_{i=1}^{|\mathcal{V}|} \exp(z_i(t))}, \quad \forall v \in \mathcal{V}. \label{eq:local_process}
\end{align}
The SLM then samples a draft token $d$ from this distribution, i.e., $d \sim \mathbf{x}(t)$.

\noindent \textbf{Step 2: LLM's Verification \& Correction.} \quad
The LLM also processes $\mathbf{s}(t)$ to produce a logit vector, which is then normalized to yield its own vocabulary distribution: $\mathbf{y}(t) = [y_1(t), y_2(t), \ldots, y_{|\mathcal{V}|}(t)]^\top$. It then compares the probabilities assigned to the draft token $d \in \mathcal{V}$ under both the SLM and LLM distributions. The verification and refinement proceeds as follows:
\begin{itemize}
    \item \textbf{Case 1 (Deterministic Acceptance):} If $y_d(t) \geq x_d(t)$, the draft token is accepted by the LLM.
    \item \textbf{Case 2-1 (Probabilistic Acceptance):} If $y_d(t) < x_d(t)$, the draft token is accepted with probability $y_d(t)/x_d(t)$.
    \item \textbf{Case 2-2 (Reject \& Resampling):} If $y_d(t) < x_d(t)$ and the draft token is rejected (with probability $1 - y_d(t)/x_d(t)$), the target token $d^\ast$ is sampled from the resampling distribution $\bm{p}(t) = [p_1(t), p_2(t), \ldots, p_{|\mathcal{V}|}(t)]^\top$, i.e., $d^\ast\sim\bm{p}(t)$, where:
    \begin{equation}
    p_v(t) = \frac{(y_v(t) - x_v(t))^+}{\sum_{i=1}^{|\mathcal{V}|} (y_i(t) - x_i(t))^+}, \quad \forall v \in \mathcal{V}, \label{resampling}
    \end{equation}
    in which $(\cdot)^+ = \max(0, \cdot)$ is the ReLU function.
\end{itemize}
The final response token $r(t)$ is set to $d$ in \textbf{Case 1} and \textbf{Case 2-1}, where the draft token is accepted, and to $d^\ast$ in \textbf{Case 2-2}, where resampling occurs. This completes one round of token generation in the HLM framework.

This mechanism is inspired by speculative decoding methods~\cite{leviathan2023fast, chen2023accelerating}, and is carefully constructed to satisfy an \textit{unbiasedness condition}, ensuring that the overall output distribution of HLM inference matches that of the original LLM:
\begin{equation}
x_v(t) \cdot (1 - \beta_v(t)) + \left(\sum_i \left( x_i(t) \cdot \beta_i(t) \right)\right) \cdot p_v(t) = y_v(t), \quad \forall v, \label{eq:equality}
\end{equation}
where $\beta_v(t) = \left(1- \frac{y_v(t)}{x_v(t)}\right)^+$ denotes the rejection probability of the $v$-th token. 

Each term in \eqref{eq:equality} has a clear interpretation, directly corresponding to the three possible cases in the LLM verification process. The first term, $x_v(t) \cdot (1 - \beta_v(t))$, represents the probability that token $v$ is accepted as the response token—either deterministically as in \textbf{Case 1} or probabilistically as in \textbf{Case 2-1}. The second term, $(\sum_i \left( x_i(t) \cdot \beta_i(t) \right)) \cdot p_v(t)$, accounts for the overall probability that token $v$ is selected through resampling in \textbf{Case 2-2}, averaged over all possible draft tokens that may be rejected. The right-hand side, $y_v(t)$, denotes the target probability assigned to token $v$ by the full LLM inference. This equality serves as a core theoretical condition that ensures the inference accuracy of the HLM framework. \tblue{To quantify deviations from this condition in later extensions, we define the \emph{bias} at round $t$ as}
\begin{align} \mathsf{Bias}(t) \!\!=\!\! \sum_{v\in\mathcal{V}}\! \left| x_v(t) (1 - \beta_v(t)) \!\!+\!\! \Big(\sum_{i \in \mathcal{V}} x_i(t) \beta_i(t)\Big) q_v(t) \!-\! y_v(t) \right|. \end{align}
\tblue{Under the vanilla HLM design, \eqref{eq:equality} implies $\mathsf{Bias}(t)=0$ for all $t$.
This bias formulation will serve as a reference metric when introducing uncertainty-aware skipping and vocabulary compression in subsequent sections.}

To proceed to the next round, the response sequence is updated in an autoregressive manner: $\mathbf{r}(t) = \mathbf{r}(t{-}1) \oplus r(t),$
so that the newly generated token becomes part of the input for the next step. This iterative procedure continues until the sequence reaches its maximum length $|\mathbf{r}(t)| = r_{\text{max}}$ or an End-of-Sentence (EOS) token is generated.

\vspace{-10pt}
\subsection{Wireless Communication}
The HLM inference process requires both uplink and downlink transmissions between the device and the BS. For the uplink, after the SLM generates a draft token, the device transmits: 1) the index of the draft token $d$ within the vocabulary, and 2) a set of indices and corresponding probabilities from the vocabulary distribution $\mathbf{x}(t)$ of \eqref{eq:local_process}, enabling the LLM to perform verification and, if necessary, resampling. Subsequently, the device downloads the index of the final response token—either the accepted draft token $d$ or a resampled target token $d^*$.

For simplicity, we assume that the communication cost incurred by transmitting token indices is negligible. This allows us to account only for the uplink transmission of the vocabulary distribution, whose payload size \( B \) is given by:
\begin{equation} \label{eq:payload}
    B = |\mathcal{V}| (b_{\text{prob}} + b_{\text{index}}) \text{ (in bits)},
\end{equation}
where \( b_{\text{prob}} \) and \( b_{\text{index}} \) denote the number of bits required to represent a probability value and a vocabulary index, respectively. For indices, binary encoding is assumed, yielding \( b_{\text{index}} = \left\lceil \log_2 |\mathcal{V}| \right\rceil \).

We adopt a block fading model for the wireless channel, where the channel gain remains constant within each round but may vary from one round to another. The uplink transmission time for the \( t \)-th round, denoted by \( \tau(t) \), is calculated using Shannon’s capacity formula as:
\begin{equation} \label{SNR}
    \tau(t) = \frac{B}{W \log_2{(1 + \text{SNR}(t))}} \text{ (in sec)},
\end{equation}
where \( W \) represents the bandwidth of the uplink channel. The signal-to-noise ratio (SNR) at the \( t \)-th round, expressed as \( \text{SNR}(t) \), depends on the transmit power \( p \) of the device, the distance \( \rho \) to the BS, the path loss exponent \( \alpha \), and the noise power \( N \).

 \subsection{Token Throughput}
To quantify the token generation rate of the HLM while accounting for both communication and computation latency, we introduce the notion of \textit{token throughput}, defined as the number of response tokens generated per unit time. For tractability, we assume that the SLM and LLM incur constant per-token computation times, denoted by \(\tau_{\text{SLM}}\) and \(\tau_{\text{LLM}}\), respectively. Then, the token throughput at the \(t\)-th round, denoted by \( \mathsf{TP}(t) \), is given by:
\begin{equation}
\mathsf{TP}(t) = 
\frac{1}{\tau_{\text{SLM}} + \tau(t) + \tau_{\text{LLM}}} \quad \text{(tokens/sec)}.
\label{eq:spec_latency}
\end{equation}

Despite its architectural advantages, HLM inference suffers from a fundamental limitation—its low token throughput, as quantified in~\eqref{eq:spec_latency}.  
Each token generation involves not only computation on both the SLM and LLM but also significant communication overhead, including the transmission of a payload proportional to the full vocabulary distribution size as defined in~\eqref{eq:payload}.  
According to~\eqref{eq:spec_latency}, increasing token throughput requires reducing the overall latency, particularly the LLM computation latency \(\tau_{\text{LLM}}\) and the uplink communication latency \(\tau(t)\). The following schemes address this by improving transmit opportunity and reducing uplink communication cost, respectively.

\vspace{-10pt}
\section{Uncertainty-Aware Opportunistic \\Hybrid Language Model} \label{sec:U-HLM}
In this section, we introduce the \textit{Uncertainty-Aware Opportunistic HLM (U-HLM)}. \tblue{U-HLM extends the standard HLM inference pipeline by enabling the device to opportunistically skip uplink transmission and LLM-side verification based on on-device uncertainty estimates. By selectively bypassing communication and computation for reliable draft tokens, U-HLM improves communication efficiency while maintaining inference accuracy.}

\vspace{-10pt}
\subsection{\tblue{Overview}} \label{sec:uhlm_overview}

\tblue{U-HLM is built upon a simple but critical observation: 
in standard HLM inference, many draft tokens generated by the device-side SLM are ultimately accepted by the server-side LLM. 
If such tokens can be reliably identified in advance, their uplink transmission and LLM verification become unnecessary.}

\tblue{The central design challenge is therefore prediction. 
Specifically, we ask whether the LLM’s rejection probability for a draft token can be inferred using only information available on the device.}

\tblue{Our key hypothesis is that the SLM’s on-device uncertainty provides a predictive signal of the LLM’s rejection behavior. 
Here, uncertainty is defined as a quantitative measure of the SLM’s confidence in its draft token, 
estimated via the consistency of its predictions under controlled inference-time perturbations. 
Since uncertainty can be computed entirely on the device prior to transmission, 
it serves as a practical proxy for the latent rejection probability at the server.}

\tblue{If this hypothesis holds, draft tokens with sufficiently low predicted rejection probability can be safely skipped, 
thereby reducing communication and computation overhead without compromising inference accuracy.}

\tblue{The remainder of this section develops U-HLM from operation to analysis. 
Section~\ref{sec:uhlm_step} presents the step-by-step inference procedure. 
Section~\ref{sec:uhlm_uncertainty} empirically validates the uncertainty–rejection relationship. 
Finally, Section~\ref{section4b} derives an uncertainty threshold that balances communication efficiency and rejection risk.}
The SLM, LLM, and datasets used are consistent with those described in Section~\ref{sec:eval}.

\vspace{-10pt}
\subsection{\tblue{Step-by-Step Operation of U-HLM.}} \label{sec:uhlm_step}

\begin{figure}[t]
\centering
\includegraphics[width=\linewidth]{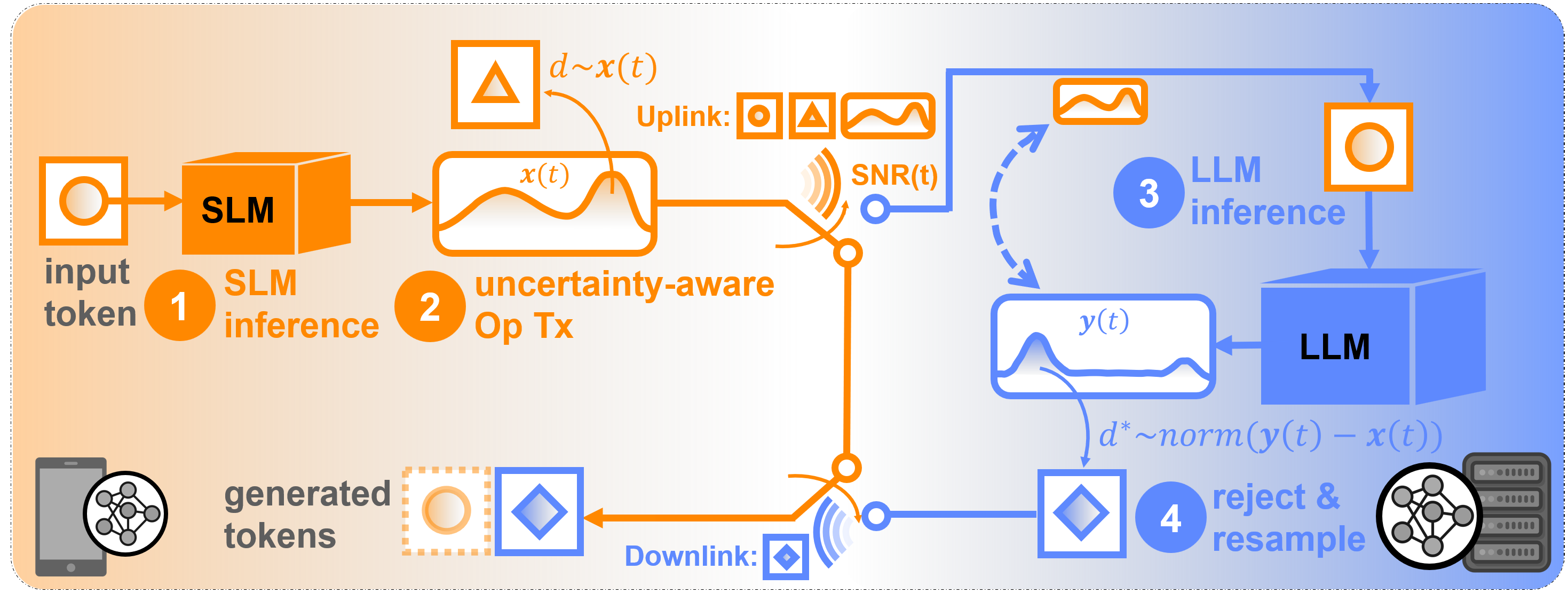}
\caption{\tblue{Schematic illustration of U-HLM framework.}}
\label{fig:11_1}
\vspace{-10pt}
\end{figure}

\tblue{
U-HLM follows the same autoregressive token generation paradigm as the standard HLM,
but introduces an additional uncertainty-aware decision stage before uplink transmission.
Fig.~\ref{fig:11_1} visualizes this process.
Below, we describe one inference round step by step.
}

\noindent \tblue{\textbf{Step 1: Draft Token Generation at the Device.}} \quad
\tblue{
At round $t$, the device-side SLM generates a draft token $d$ from its vocabulary distribution $\mathbf{x}(t)$,
conditioned on the current input sequence.
This step is identical to the draft generation stage in standard HLM inference.
}

\noindent \tblue{\textbf{Step 2: On-Device Uncertainty Estimation via Temperature Perturbation.}} \quad
\tblue{
After generating the draft token $d$ at round $t$, the device evaluates the reliability of this draft
by estimating its uncertainty using temperature perturbation \cite{gao2024spuq}.
Specifically, the device samples $M$ temperature values
$\{\theta^{(1)}, \ldots, \theta^{(M)}\}$ from the interval $[0, \theta_{\max}]$. In our experiments, we set $M = 20$ and $\theta_{\max} = 2$, with temperatures sampled uniformly from this interval. For each temperature $\theta^{(m)}$, the SLM re-normalizes its logits through a temperature-scaled
softmax operation, producing a perturbed vocabulary distribution
$\tilde{\mathbf{x}}^{(m)}(t)$, whose $v$-th element is given by
}
\begin{align}
\tblue{
\tilde{x}_v^{(m)}(t) = 
\frac{\exp(z_v(t)/\theta^{(m)})}
{\sum_{i=1}^{|\mathcal{V}|} \exp(z_i(t)/\theta^{(m)})},
\quad \forall v \in \mathcal{V}.
}
\label{eq:temp_perturb}
\end{align}
\tblue{
From each perturbed distribution, a token $d^{(m)}$ is sampled.
The uncertainty $u(t)$ is then quantified as the average disagreement
between these sampled tokens and the original draft token $d$:
}
\begin{equation}
\tblue{
u(t) = \frac{1}{M} \sum_{m=1}^{M} \mathds{1}(d^{(m)} \neq d),
}
\label{eq:uncertainty}
\end{equation}
\tblue{
where $\mathds{1}(\cdot)$ denotes the indicator function}\footnote{Regarding the concern that uncertainty estimation may increase the SLM's computation overhead, we note that it is fully parallelized with the standard forward pass and thus does not introduce additional latency.}.

\noindent \tblue{\textbf{Step 3: Uncertainty-Aware Skipping Decision.}} \quad
\tblue{
Based on the estimated uncertainty, the device determines whether to trigger uplink transmission
and LLM-side verification.
To formalize this decision, we introduce a binary indicator variable
$\delta(t) \in \{0,1\}$, where $\delta(t)=1$ indicates that uplink transmission
and LLM computation are performed at round $t$, and $\delta(t)=0$ otherwise.
The skipping decision is governed by a fixed uncertainty threshold $u_{\mathrm{th}}$:
}
\begin{equation}
\tblue{
\delta(t) =
\begin{cases}
0, & \text{if } u(t) \leq u_{\mathrm{th}}, \\
1, & \text{otherwise}.
\end{cases}
}
\label{eq:u_ops}
\end{equation}
\tblue{
When $\delta(t)=0$, the draft token is directly appended to the response sequence without uplink transmission or LLM verification.
When $\delta(t)=1$, the device transmits the draft token together with the SLM’s vocabulary distribution $\mathbf{x}(t)$ to the BS.
}

\noindent \tblue{\textbf{Step 4: Conditional LLM Verification and Correction.}} \quad
\tblue{
When $\delta(t)=1$, the BS-side LLM first performs inference to obtain its vocabulary distribution $\mathbf{y}(t)$.
It then applies the standard HLM verification procedure by comparing the draft probability under $\mathbf{x}(t)$ and $\mathbf{y}(t)$,
either accepting the draft token or replacing it through resampling from the distribution defined in~\eqref{resampling}.
}

\noindent \tblue{\textbf{Step 5: Sequence Update and Synchronization.}} \quad
\tblue{
The corrected token is appended to the response sequence, and inference proceeds to the next round.
When skipping occurs ($\delta(t)=0$), the device and BS may temporarily diverge in their local token indices.
To maintain consistency, a lightweight index-based resynchronization is performed,
which incurs negligible communication overhead and is therefore omitted from the cost analysis.
}

\tblue{
By conditionally bypassing uplink transmission and LLM computation,
U-HLM reduces communication and computation costs while preserving inference accuracy
through uncertainty-aware control.
}

\vspace{-10pt}
\subsection{Uncertainty and Rejection Prediction} \label{sec:uhlm_uncertainty}

\tblue{Since U-HLM relies on uncertainty as a proxy for rejection probability, 
it is essential to examine whether such a relationship holds in practice. 
We empirically analyze the correlation between on-device uncertainty and the LLM’s rejection probability, 
and compare multiple uncertainty estimation methods to identify the most predictive metric.}

We conduct experiments using representative uncertainty estimation techniques, including (1) sample-based methods \cite{huang2023look} (Bayesian methods such as MC Dropout \cite{gal2016dropout}) and (2) perturbation-based methods \cite{gao2024spuq} (test-time augmentation methods such as prompt perturbation and temperature perturbation). In the MC Dropout setting, we apply dropout to the SLM using $20$ dropout probabilities sampled uniformly from the range \([0, 0.1]\), sampling one token per dropout configuration. For prompt perturbation, we generate $20$ paraphrased versions of the input token sequence using WordNet \cite{miller1995wordnet} and sample a token for each paraphrased prompt. For temperature perturbation, we sample $20$ temperatures uniformly from \([0, 2]\), generating one token per temperature. \tblue{The uncertainty for each method is quantified as the average disagreement between the generated tokens and the draft token, with the temperature perturbation case formally defined in Eq.~\eqref{eq:uncertainty}.}


\begin{figure}[t]
\centering
\includegraphics[width=\linewidth]{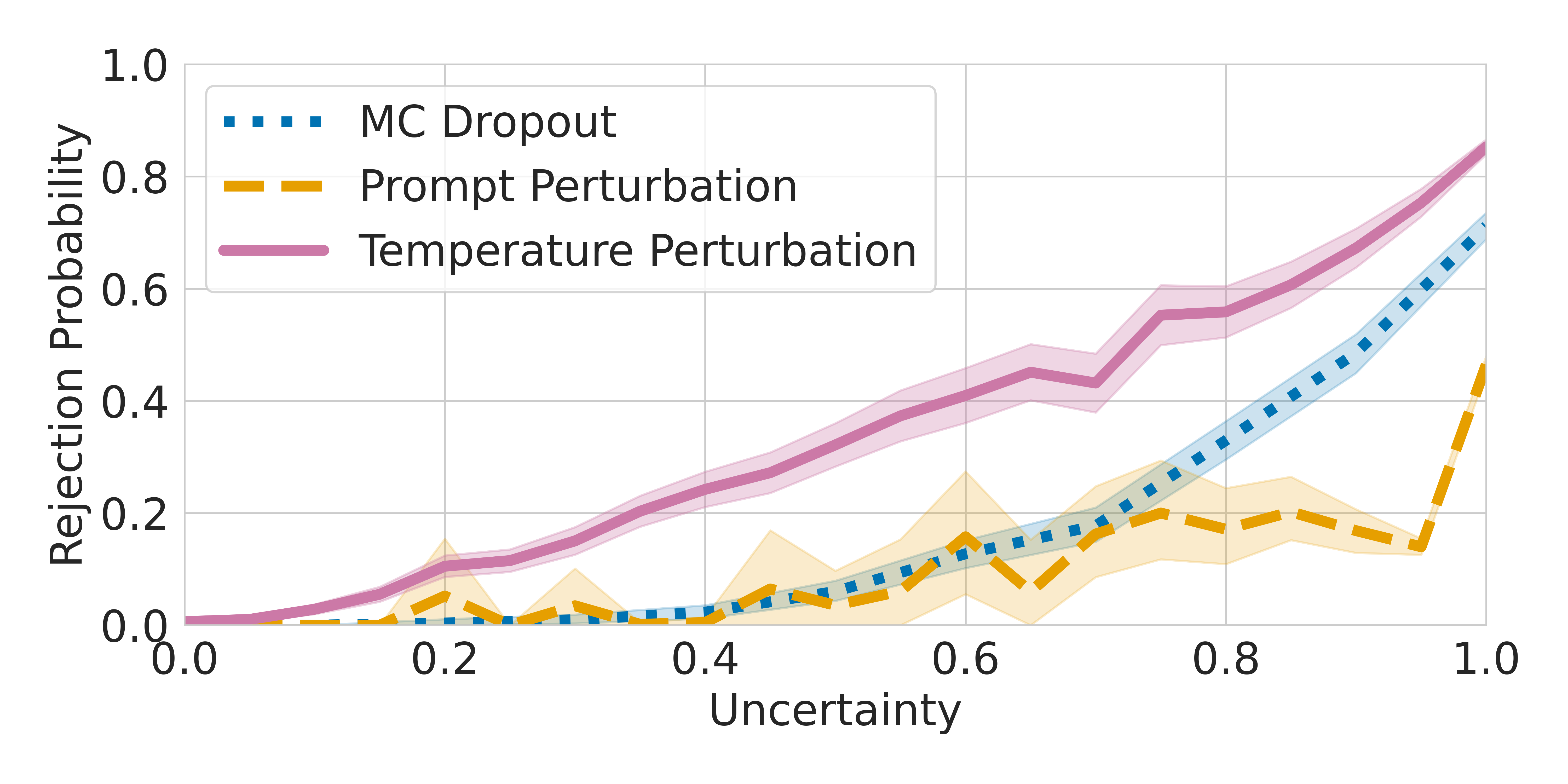}
\caption{Uncertainty vs. rejection probability. Solid lines show mean; shaded regions denote 95\% CI. Correlation coefficients: temperature perturbation (0.71), MC Dropout (0.65), and prompt perturbation (0.20).}
\label{fig:cnnvit_X}
\vspace{-10pt}
\end{figure}

\figref{fig:cnnvit_X} shows the correlation between the uncertainty and rejection probability across these methods, leading to the following observations:
\begin{itemize}
    \item All three uncertainty measures exhibit a positive correlation with the rejection probability.
    \item Temperature perturbation demonstrates the clearest linear trend across a wide range of uncertainty values and yields the highest correlation coefficient (0.7106).
\end{itemize}
These results suggest that temperature perturbation is the most effective approach for predicting rejection probability among the techniques we concern. Motivated by this strong empirical linear trend, we approximate the rejection probability as a linear function of uncertainty. Accordingly, we formalize this relationship as follows:
\begin{remark}[Linear Correlation Between Uncertainty and Rejection Probability] \label{remark:1}
For every $t$-th round in HLM inference, the uncertainty \(u(t)\) measured via temperature perturbation exhibits a linear relationship with the LLM's rejection probability \(\beta_d(t)\) for the SLM-generated draft token \(d\) as:
\begin{equation}
    \beta_d(t) = a \cdot u(t) + b, \quad \forall t, \label{eq:uncertainty_rejection2}
\end{equation}
where \(a = 0.815\) and \(b = -0.066\) are obtained through linear regression, with a mean squared error (MSE) of $1.41\times 10^{-3}$ and a coefficient of determination of \(0.9774\).
\end{remark}

\vspace{-10pt}
\subsection{Uncertainty Threshold} \label{section4b}
This subsection focuses on designing the uncertainty threshold to maximize communication efficiency while ensuring that the inference degradation remains bounded. In the context of U-HLM, inference degradation occurs when a draft token that would have been rejected by the LLM is mistakenly skipped—i.e., a false positive. We define the probability of such events as the U-HLM’s rejection risk \( R \), which exhibits a fundamental trade-off with communication efficiency, governed by the choice of uncertainty threshold. Accordingly, our goal is to determine an uncertainty threshold to minimize communication cost without allowing the rejection risk \( R \) to exceed a tolerable bound.

Recall that in the HLM framework, a draft token \(d\) is rejected when \(y_d(t) < x_d(t)\), with a rejection probability given by \(1 - y_d(t)/x_d(t)\). From the device's perspective, \(x_d(t)\) is fixed once \(d\) is selected, while \(y_d(t)\) remains a latent and inaccessible random variable. Consequently, the expected rejection probability \(\mathbb{E}[\beta_d(t)]\) conditioned on a fixed draft token probability \(x_d(t)\) can be expressed as:
\begin{align}
\mathbb{E}_{y_d(t)}[\beta_d(t)] 
\!=\!\! P(y_d(t) \!<\! x_d(t)) 
 \mathbb{E}_{y_d(t)}\!\!\left[ 1\!-\!\frac{y_d(t)}{x_d(t)} \!\mid\! y_d(t) \!<\! x_d(t) \!\right]\!\!. \label{eq:10}
\end{align}
Given that the correlation between uncertainty and rejection probability exhibits only minor deviations (see \figref{fig:cnnvit_X}), we approximate \(\mathbb{E}[\beta_d(t)] \approx \beta_d(t)\), which is now able to leverage the linear relationship in~\eqref{eq:uncertainty_rejection2}.

\begin{figure}[t]
    \centering
    \includegraphics[width=\linewidth]{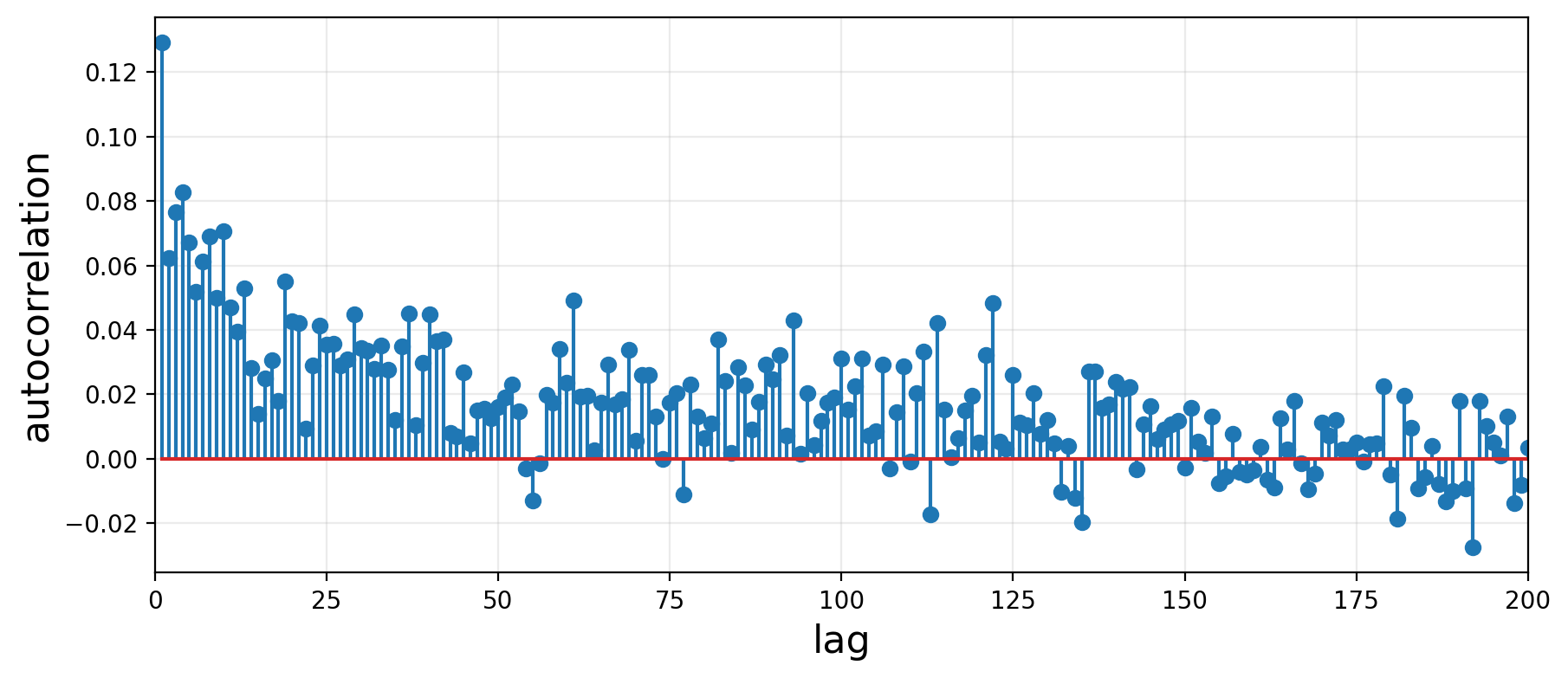}
    \caption{\tblue{Token-level autocorrelation of uncertainty across generation steps as a function of lag.}}
    \label{fig:token_acf}
    \vspace{-10pt}
\end{figure}

\tblue{For analytical tractability, we assume that the uncertainty process $u(t)$ is independent and identically distributed (i.i.d.) over time.
The rejection probability $\beta_d(t)$ is then deterministically related to $u(t)$ through the linear mapping in~\eqref{eq:uncertainty_rejection2}.}
\begin{remark}[\tblue{Uncertainty Independence Validation}]
\tblue{Although language generation is autoregressive, we empirically examine the temporal dependence of token-level uncertainty. Fig.~\ref{fig:token_acf} shows the autocorrelation of uncertainty across generation steps.
The maximum absolute autocorrelation is 0.13 at lag~1, and rapidly decays to values below 0.1 for larger lags.
This indicates weak and short-range dependence without persistent temporal correlation.
Therefore, adopting an approximate independence assumption for analytical tractability does not significantly deviate from observed behavior in practice.}
\end{remark}

Under this assumption, a fixed uncertainty threshold can be derived uniformly across rounds, as formalized in the following result.
\begin{theorem}[Uncertainty Threshold and Rejection Risk] \label{theorem:combined} 
Assuming i.i.d. uncertainty \(u(t)\) and rejection probability \(\beta_d(t)\), let \(u \coloneqq u(t)\) and \(\beta \coloneqq \beta_d(t)\) for any \(t\). Besides, we regard $\beta$ equivalent to its expectation $\mathbb{E}[\beta]$ of \eqref{eq:10}. Then, the uncertainty threshold \(u_{\text{th}}\) is given as the upper limit:
\begin{equation}
u_{\text{th}} = \frac{\Delta - b}{a},
\label{Eq:skip}
\end{equation}
where \(\Delta = P(y_d < x_d)\) denotes the probability that a token is not deterministically accepted.
The resultant rejection risk \(R\) when skipping tokens with \(u \leq u_{\text{th}}\) is upper bounded~as:
\begin{equation}
R \leq \frac{\Delta^{3/2}}{\sqrt{3a}} \cdot \sqrt{\int_{u=-\frac{b}{a}}^{\frac{\Delta - b}{a}} |f(u)|^2 \, du},
\label{upperbound}
\end{equation}
where \(f(u)\) denotes the probability density function (PDF) of the uncertainty \(u\).
\vspace{-10pt}
\begin{proof}
Since \(0 < \mathbb{E}_{y_d}\left[ 1-\frac{y_d}{x_d} \mid y_d < x_d \right] \leq 1\), it follows from \eqref{eq:10} that \(0 < \beta \leq \Delta\). Using the linear relationship in \eqref{eq:uncertainty_rejection2}, the uncertainty range can be bounded as:
\begin{equation}
-\frac{b}{a} < u \leq \underbrace{\frac{\Delta - b}{a}}_{:= u_{\text{th}}}.
\label{bounds}
\end{equation}
Defining $R \coloneqq \int_{u=-\frac{b}{a}}^{\frac{\Delta - b}{a}} (au + b) \cdot f(u) \, du$ and applying Hölder's inequality to the integral yields the upper bound on \(R\).
\end{proof}
\end{theorem}
\vspace{-7pt}
As shown in \eqref{bounds}, we refer to setting the threshold at the upper bound as risk-prone skipping, and to using the lower bound as risk-averse skipping. The rejection risk \( R \) in \textbf{Theorem}~\ref{theorem:combined} quantifies the expected increase in rejection probability induced by risk-prone skipping, as determined by the uncertainty threshold and the density of the uncertainty distribution. This effectively captures the potential degradation in inference accuracy due to bypassing uplink transmission for tokens that may ultimately be rejected.

\begin{figure}[t]
\centering
\begin{minipage}{0.48\textwidth}
    \centering
    \includegraphics[width=\textwidth]{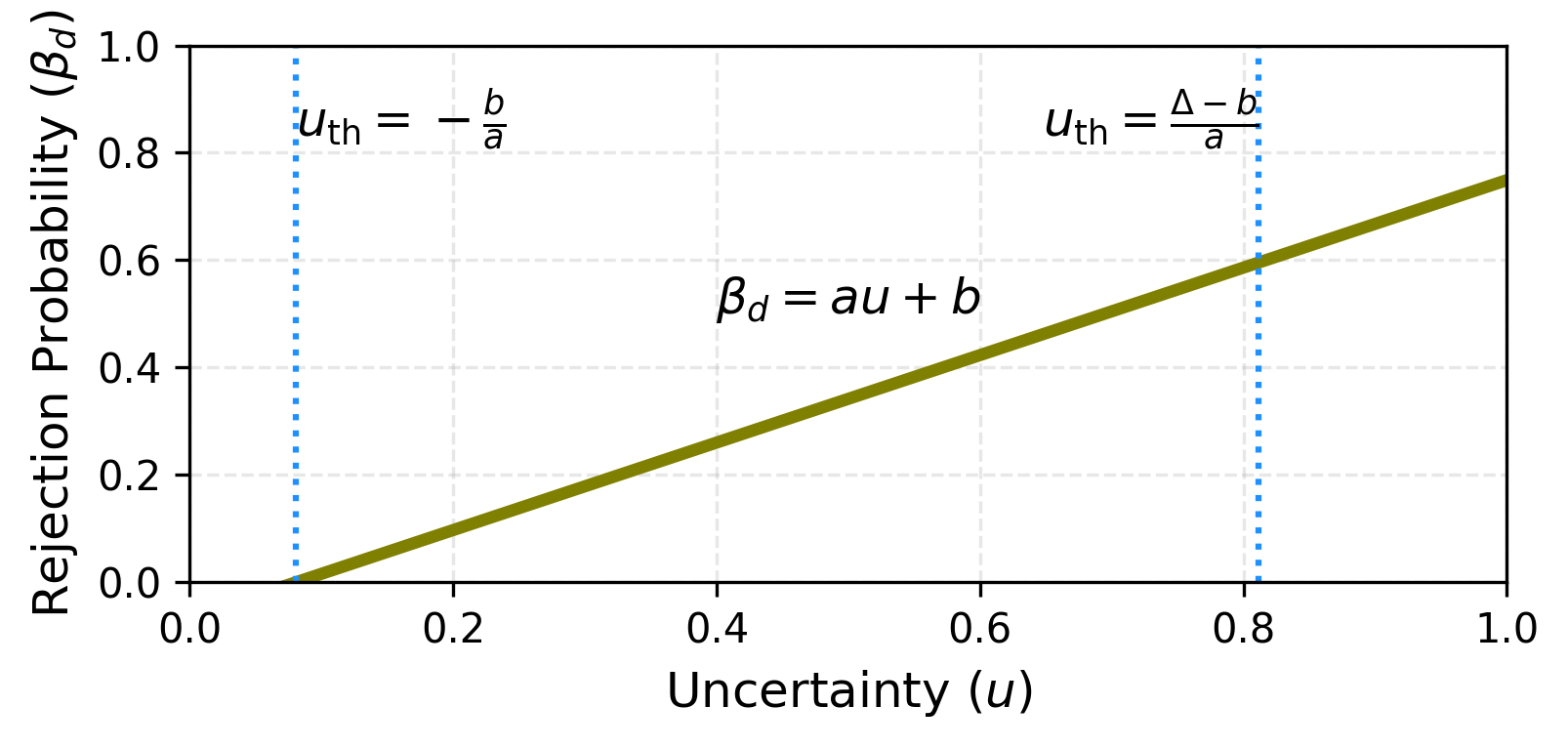}
    \subcaption{Linear relationship between uncertainty and rejection probability. Dashed vertical lines indicate the theoretical risk-averse and risk-prone thresholds, respectively.}
    \label{fig:u_21}
\end{minipage}
\hfill
\begin{minipage}{0.48\textwidth}
    \centering
    \includegraphics[width=\textwidth]{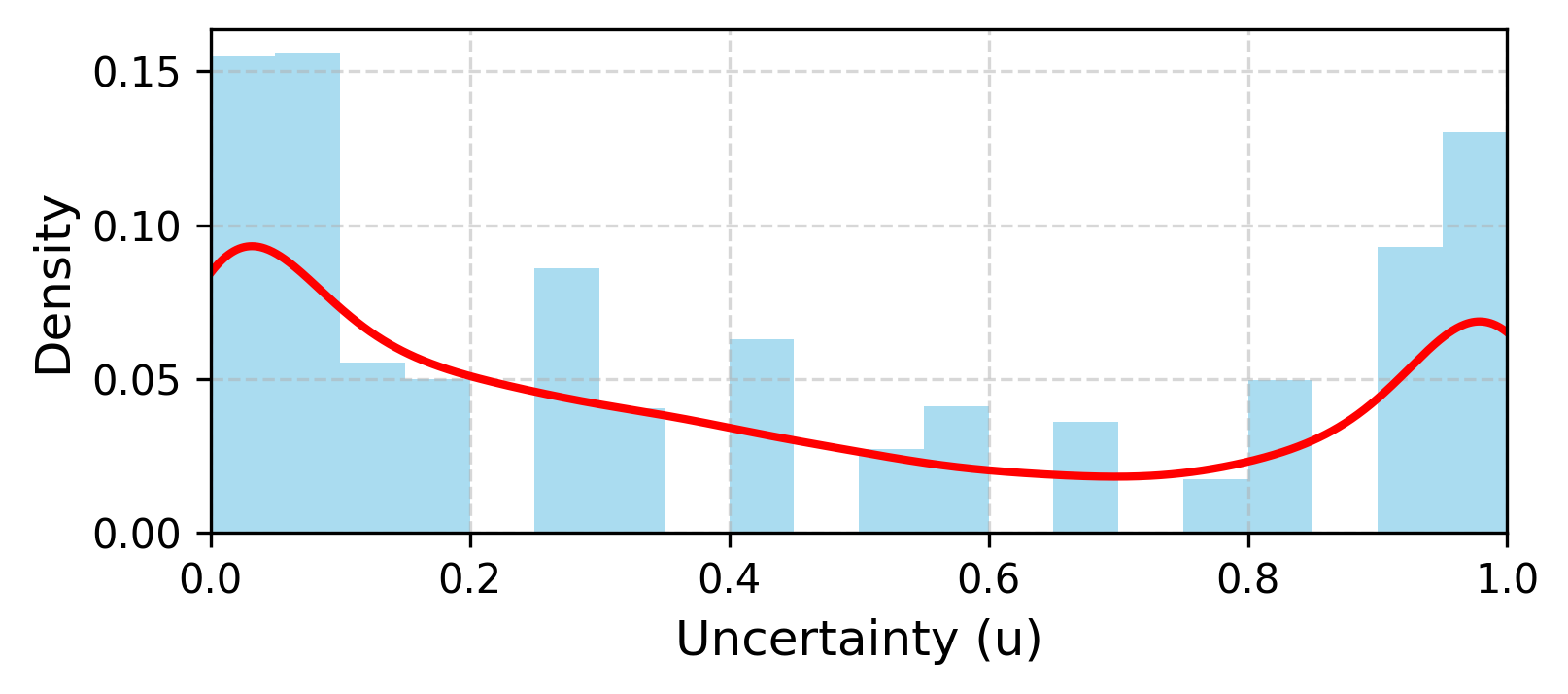}
    \subcaption{Probability Density Function (PDF) of uncertainty, obtained via Gaussian kernel density estimation (KDE).}
    \label{fig:u_11}
\end{minipage}
\caption{Empirical characterization of two uncertainty thresholds and the density of the uncertainty values.}
\label{fig:combined12}
\vspace{-10pt}
\end{figure}

\figref{fig:combined12} presents an empirical characterization of the theoretically derived uncertainty thresholds.  
Given \(\Delta = 0.5956\) under our simulation setup, the thresholds computed from Theorem~\ref{theorem:combined} are \(\frac{\Delta - b}{a} = 0.8117\) for risk-prone skipping and \(-\frac{b}{a} = 0.0810\) for risk-averse skipping.  
These thresholds are visually identified in \figref{fig:u_21}, where the risk-averse threshold corresponds to the point at which the rejection probability first reaches zero—indicating that skipping occurs only for deterministically accepted tokens. Moreover, based on the empirical uncertainty distribution shown in \figref{fig:u_11}, the expected rejection risk is measured as \(R = 4.94 \times 10^{-3}\), which satisfies the theoretical upper bound \(R < 9.87 \times 10^{-3}\).  
The upper bound in \eqref{upperbound} also suggests that a sharper slope \(a\), induced by temperature perturbation, leads to a tighter rejection risk bound.

\begin{figure}[t]
    \centering
    \subfloat[\tblue{Position-wise bias statistics.}]{
        \includegraphics[width=0.47\textwidth]{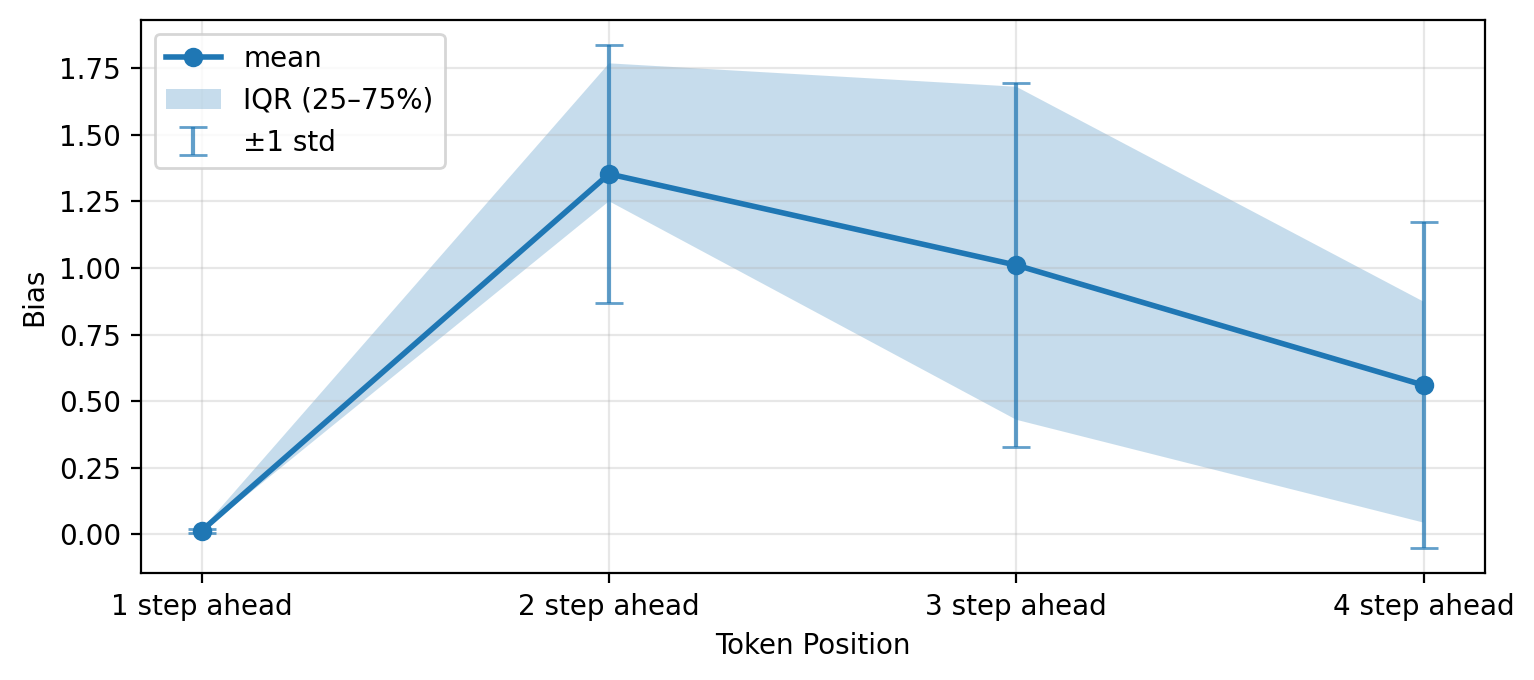}
        \label{fig:token_position_bias}
    }
    \hfill
    \subfloat[\tblue{Position-wise uncertainty statistics with consecutive skip probability (dashed).}]{
        \includegraphics[width=0.47\textwidth]{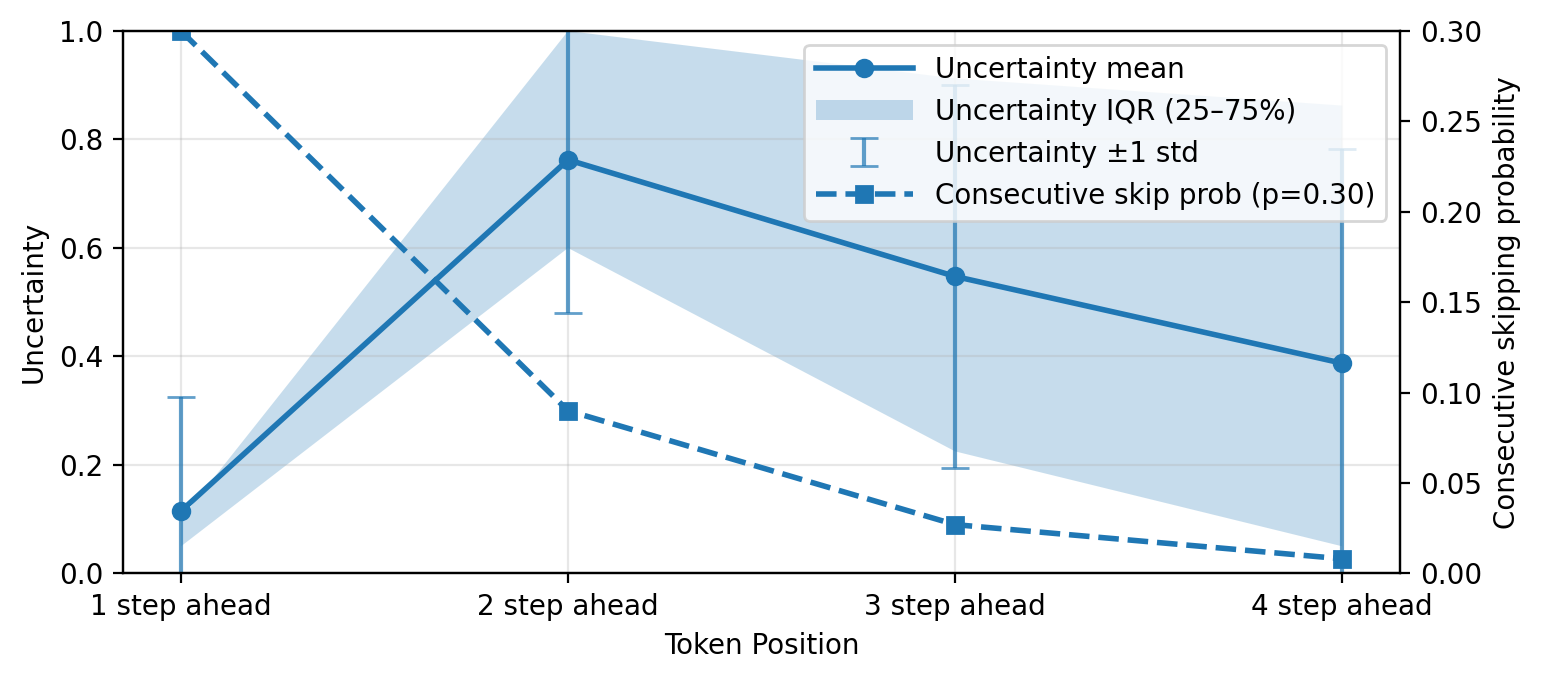}
        \label{fig:token_position_uncertainty}
    }
    \caption{
    \tblue{Token position-wise dynamics of bias and uncertainty across consecutive autoregressive steps.}
    }
    \label{fig:token_position_stats}
    \vspace{-10pt}
\end{figure}

\tblue{Fig.~\ref{fig:token_position_stats} illustrates how uncertainty and a bias proxy evolve across consecutive autoregressive steps under uncertainty-aware skipping. 
The two quantities exhibit aligned position-dependent trends, indicating that higher uncertainty is associated with larger distributional deviation. 
Since deviation from the unbiasedness identity can only arise when a token with nonzero rejection probability is finalized without LLM verification, bias is intrinsically coupled to skip events. 
Immediately after an LLM refinement step, both uncertainty and bias are minimal, indicating a reset to a low-distortion regime. 
At the subsequent autoregressive step, bias increases due to the absence of verification; however, as additional tokens accumulate, the distortion gradually decreases and stabilizes, exhibiting a rise–decay–plateau pattern rather than monotonic growth. 
This reduction is attributable to the autoregressive conditioning effect, where the expanded context progressively constrains the SLM prediction distribution and mitigates earlier local deviations.}
\vspace{-7pt}
\begin{remark}[\tblue{Sequence-level behavior under uncertainty-aware skipping}]
\label{rem:sequence_level_skipping}
\tblue{From a sequence-level perspective, cumulative distortion is primarily associated with consecutive skip events.}

\tblue{Under the calibrated risk-averse threshold configuration, the empirical single-step skipping probability remains below approximately $30\%$ (see Fig.~\ref{fig:u_3}), implying that consecutive skip runs decay geometrically (e.g., $0.3^2 \approx 0.09$). 
Although temporary bias increases are observed during such consecutive skipping events, each LLM refinement step restores the generation process to a low-distortion regime. 
As a result, the system rarely remains in elevated distortion states for prolonged periods.}

\tblue{Accordingly, while no closed-form bound on sequence-level divergence is derived, the observed autoregressive dynamics do not exhibit sustained or monotonic distortion growth across decoding steps under the evaluated configuration.}
\end{remark}
\vspace{-7pt}
\tblue{These observations provide empirical support for the theoretical design principles developed in this section.}

\begin{figure}[t]
\centering
\includegraphics[width=\linewidth]{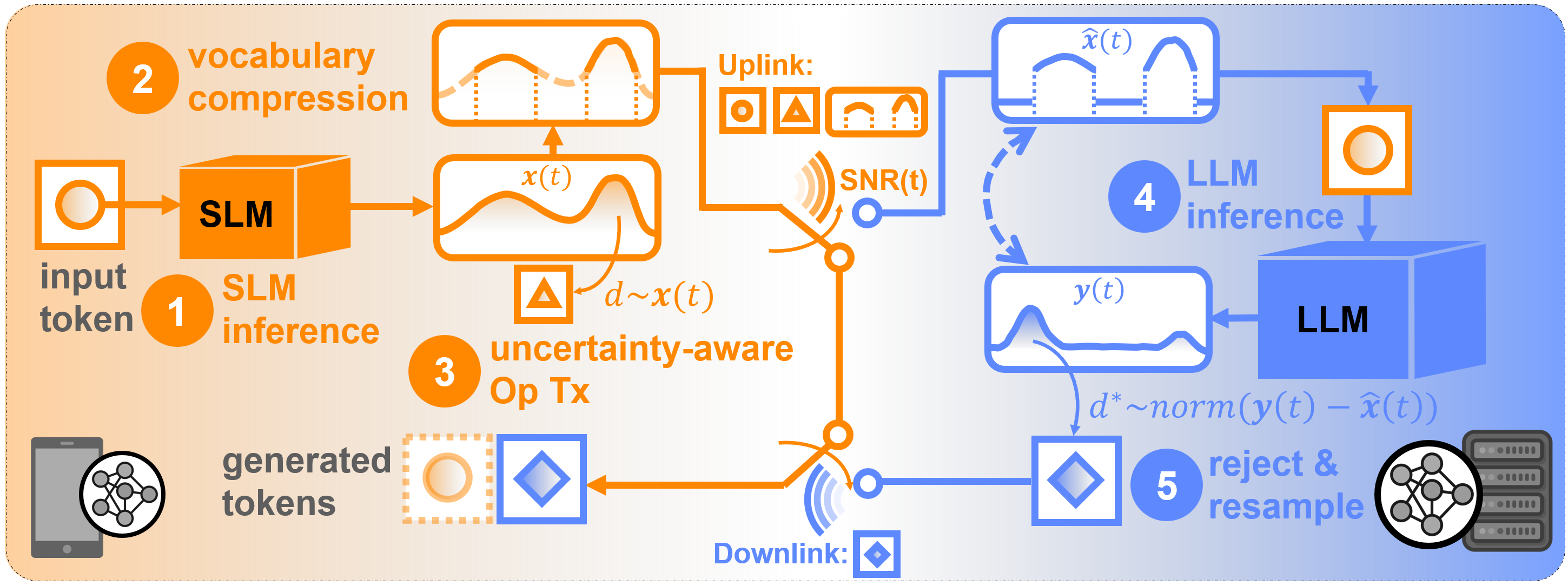}
\caption{\tblue{Schematic illustration of CU-HLM framework. CU-HLM extends the uncertainty-aware opportunistic transmission in U-HLM by incorporating compressed vocabulary transmission for high-uncertainty tokens.}}
\label{fig:11_2}
\vspace{-10pt}
\end{figure}


\section{Compressed Vocabulary Transmission for Communication-Efficient U-HLM} \label{sec:CU-HLM}

\tblue{
While U-HLM reduces communication overhead by opportunistically skipping uplink transmission for low-uncertainty tokens, 
high-uncertainty tokens still require full vocabulary transmission to the server. 
This motivates a further extension that targets communication efficiency even when skipping is not possible.
}

\tblue{
In this section, we propose \textit{Communication-Efficient U-HLM with Compressed Vocabulary Transmission (CU-HLM)}, 
which augments U-HLM by introducing vocabulary compression during required transmissions. 
Beyond merely truncating the vocabulary, CU-HLM systematically analyzes the trade-off between compression level and inference accuracy, 
and derives optimal compression strategies under both offline and online settings.
}


\subsection{\tblue{Overview}} \label{sec:CU_overview}

\tblue{
CU-HLM extends U-HLM along a complementary dimension of communication reduction. 
While U-HLM determines \emph{whether} uplink transmission is required via uncertainty-aware skipping, 
CU-HLM optimizes \emph{how much} information is transmitted when communication becomes necessary.
}

\tblue{
Specifically, when $u(t) > u_{\text{th}}$, instead of transmitting the full vocabulary distribution, 
the device sends only a compressed subset $\bar{\mathcal{V}} \subset \mathcal{V}$. 
For notational brevity, we denote its (possibly time-varying) cardinality as $k(t) = |\bar{\mathcal{V}}|$, 
which serves as a key design parameter controlling the compression level.
}

\tblue{
To enable principled compression design, we first quantify the distortion induced by vocabulary truncation using the total variation distance (TVD) between the original and distorted resampling distributions. 
We then formulate an optimization problem that determines the smallest vocabulary size $k(t)$ satisfying a bounded distortion constraint.
}

\tblue{
The remainder of this section develops CU-HLM from operation to analysis. 
Section~\ref{sec:CU_operation} presents the step-by-step operation of CU-HLM, illustrating how compressed vocabulary transmission is integrated with uncertainty-aware triggering. 
Section~\ref{sec:CU_empirical} provides empirical evidence by (i) examining the compressibility of the SLM’s vocabulary distribution and (ii) analyzing the relationship between TVD and inference bias. 
Finally, Section~\ref{Subsection: Top-k} and Section~\ref{Subsection: Top-k_v2}, respectively, derive tractable distortion bounds and yields optimal vocabulary sizes for two practical variants: 
(i) an offline compression scheme based on long-term statistical estimates, and 
(ii) an online scheme executable entirely on-device without server-side feedback.
}

\subsection{\tblue{Step-by-Step Operation of CU-HLM}} \label{sec:CU_operation}

\begin{table*}[t]
\centering
\caption{\tblue{Comparison of HLM, U-HLM, and CU-HLM in communication-aware hybrid inference.}}

\resizebox{\textwidth}{!}{%
\begin{tabular}{lccc}
\toprule
\textbf{Design Dimension} & \textbf{HLM} & \textbf{U-HLM} & \textbf{CU-HLM} \\
\midrule

Design parameters 
& -- 
& $u_{\text{th}}$ 
& $u_{\text{th}},\, k(t)$ \\

Uplink transmission 
& Full vocabulary 
& \makecell[l]{Skip if $u(t)\le u_{\text{th}}$ \\ 
Full vocab if $u(t)>u_{\text{th}}$}
& \makecell[l]{Skip if $u(t)\le u_{\text{th}}$ \\ 
Top-$k(t)$ if $u(t)>u_{\text{th}}$} \\

Model output distortion constraint
& \makecell[l]{Exact token-level unbiasedness: \\[-2pt]
$\mathrm{Bias}(t)=0$}
& \makecell[l]{Per-token rejection risk: \\[-2pt]
$R \leq \frac{\Delta^{3/2}}{\sqrt{3a}} 
\cdot \sqrt{\int_{u=-\frac{b}{a}}^{\frac{\Delta - b}{a}} 
|f(u)|^2 \, du}$}
& \makecell[l]{Token-level TVD: \\[-2pt]
$D_{\text{TV}}(\bm{p}(t),\bm{q}(t))\le\theta$} \\

Optimal design rule 
& -- 
& $u_{\text{th}}=\frac{\Delta-b}{a}$ 
& \makecell[l]{Offline: $k^*=\arg\min_k \{k\mid \mathbb{E}_t[\mathrm{U}_{\text{TV}}(k,t)]\le\theta\}$ \\ 
Online: $k(t)^*=\arg\min_{k(t)} \{k(t)\mid \overline{\mathrm{U}}_{\text{TV}}(au(t)+b)\le\theta\}$} \\

\bottomrule
\end{tabular}%
}
\label{tab:comparison}
\end{table*}

\tblue{
CU-HLM follows the same uncertainty-aware triggering mechanism as U-HLM, 
but augments the transmission stage with compressed vocabulary transmission. 
Fig.~\ref{fig:11_2} illustrates the extended framework.
In particular, Fig.~\ref{fig:11_2} highlights two key decision points:
(i) uncertainty-based triggering that determines whether uplink is needed, and
(ii) distortion-constrained top-$k(t)$ compression that controls how much information is transmitted when triggering occurs.
Below, we describe one inference round step by step, 
while the complete end-to-end procedure is summarized as pseudocode in \textbf{Algorithm}~\ref{algo_1}, 
which explicitly details the integration of uncertainty-aware triggering and compressed vocabulary transmission.
}

\noindent \tblue{\textbf{Step 1: Draft Token Generation and Uncertainty Estimation.}} \quad
\tblue{
At round $t$, the device-side SLM generates a draft token $d$ 
and estimates the uncertainty $u(t)$ as described in Section~\ref{sec:uhlm_step}.
}

\noindent \tblue{\textbf{Step 2: Uncertainty-Aware Transmission Decision.}} \quad
\tblue{
If $u(t) \le u_{\text{th}}$, opportunistic skipping is performed as in U-HLM, 
and the draft token is directly appended to the response sequence.
If $u(t) > u_{\text{th}}$, uplink transmission is triggered.
}

\noindent \tblue{\textbf{Step 3: Vocabulary Compression at the Device.}} \quad
\tblue{
When transmission is required, instead of sending the full vocabulary distribution 
$\bm{x}(t)$, the device retains the top-$k(t)$ entries of $\bm{x}(t)$ 
under descending probability ordering, forming a compressed set 
$\bar{\mathcal{V}} \subset \mathcal{V}$. 
Only the indices and probabilities of these tokens are transmitted to the server.
}

\noindent \tblue{\textbf{Step 4: Distribution Reconstruction and Verification at the Server.}} \quad
\tblue{
Upon reception, the server reconstructs an approximate vocabulary distribution 
$\hat{\bm{x}}(t)$ by preserving the transmitted probabilities 
and uniformly redistributing the residual probability mass over the remaining tokens:
}
\begin{equation}
\hat{x}_i(t) = 
\begin{cases} 
x_i(t), & \text{for } i = 1, \dots, k(t), \\
\frac{1 - \sum_{i=1}^{k(t)} x_i(t)}{|\mathcal{V}| - k(t)}, 
& \text{for } i = k(t)+1, \dots, |\mathcal{V}|.
\end{cases}
\label{eq_reconst}
\end{equation}

\tblue{
The LLM then performs the standard HLM verification procedure using 
$\hat{\bm{x}}(t)$ and its own distribution $\bm{y}(t)$.
If the draft token is rejected, resampling is performed based on the distorted resampling distribution:
}
\begin{equation}
q_v(t) = 
\frac{(y_v(t) - \hat{x}_v(t))^+}
{\sum_i (y_i(t) - \hat{x}_i(t))^+}.
\label{resampling_distortion}
\end{equation}

\noindent \tblue{\textbf{Step 5: Sequence Update and Synchronization.}} \quad
\tblue{
The accepted or resampled token is appended to the response sequence, 
and inference proceeds to the next round.
}

\tblue{
Compared to U-HLM, CU-HLM introduces additional distortion due to vocabulary truncation, 
which may affect (i) the draft token rejection probability and 
(ii) the resampling distribution. 
To focus solely on the distortion introduced in the resampling stage, 
we assume that the draft token’s index and probability are always transmitted. 
Under this assumption, the LLM’s acceptance decision for the draft token remains identical to that of the uncompressed case, 
and any performance degradation arises exclusively from the distorted resampling distribution.
}

\tblue{
For clarity, Table~\ref{tab:comparison} summarizes the key design differences among HLM, U-HLM, and CU-HLM,
highlighting their transmission policies, model output distortion constraints, and optimization strategies.
The ``Model output distortion constraint'' row reports the primary distributional control mechanism used in each design
(i.e., exact token-level unbiasedness for HLM, rejection-risk control for U-HLM, and token-level TVD control for CU-HLM).
These mechanisms are not identical, and the table is intended to clarify their respective roles rather than imply direct equivalence.
}

\begin{algorithm}[t]
\caption{Operational flow of CU-HLM.}
\label{algo_1}
\begin{algorithmic}[1]
\Require Input token sequence $\mathbf{s}$, $u_{\text{th}}$, $k(t)^*$ $\forall t$, $r_{\text{max}}$, $\theta_{\text{max}}$
\While{$|\mathbf{r}(t{-}1)| < r_{\text{max}}$ \textbf{and} $r(t{-}1)\neq\text{EOS}$}
    \State \textbf{// SLM Operation at Device (Round $t$)}
    \State Generate input $\mathbf{s}(t) = \mathbf{s} \oplus \mathbf{r}(t{-}1)$
    \State Sample $\{\theta^{(1)}, \ldots, \theta^{(M)}\}$ from $[0, \theta_{\text{max}}]$
    \State Generate vocabulary distributions $\mathbf{x}(t)$, $\tilde{\mathbf{x}}^{(m)}(t)$ $\forall m$
    \State Sample draft token $d$ and perturbed tokens $d^{(m)}$ $\forall m$
    \State Measure uncertainty $u(t)$ via \eqref{eq:uncertainty}
    \State Set $r(t) \gets d$
    \If{$u(t) > u_{\text{th}}$}
        \State Upload top $k(t)^*$ tokens $\{x_1(t), \dots, x_{k(t)^*}(t)\}$ and draft token $d$ to the BS
        \State \textbf{// LLM Operation at BS (Round $t$)}
        \State Reconstruct full distribution $\hat{\mathbf{x}}(t)$ via \eqref{eq_reconst}
        \State Generate LLM distribution $\mathbf{y}(t)$ by processing $\mathbf{s}(t)$
        \If{$y_d(t) < x_d(t)$}
            \State With probability $1 - \frac{y_d(t)}{x_d(t)}$, sample 
 a target token $r(t) = d^\ast$ from:
            \[
            P(r(t) = d^\ast)=\frac{(y_{d^\ast}(t) - \hat{x}_{d^\ast}(t))^+}{\sum_{i=1}^{|\mathcal{V}|} (y_i(t) - \hat{x}_i(t))^+}
            \]
        \EndIf
        \State Send $r(t)$ to the device
        \ElsIf{$u(t) \leq u_{\text{th}}$}
        \State \textbf{Opportunistic Skipping}
        \EndIf
        \State Update $\mathbf{r}(t) = \mathbf{r}(t{-}1) \oplus r(t)$, $t \gets t + 1$
\EndWhile
\end{algorithmic}
\end{algorithm}

\subsection{\tblue{Vocabulary Compression: Empirical Foundations}}
\label{sec:CU_empirical}

\subsubsection{\tblue{Vocabulary Compressibility Analysis}}

\begin{figure}[t]
\centering
\begin{subfigure}[t]{\linewidth}
\includegraphics[width=\linewidth]{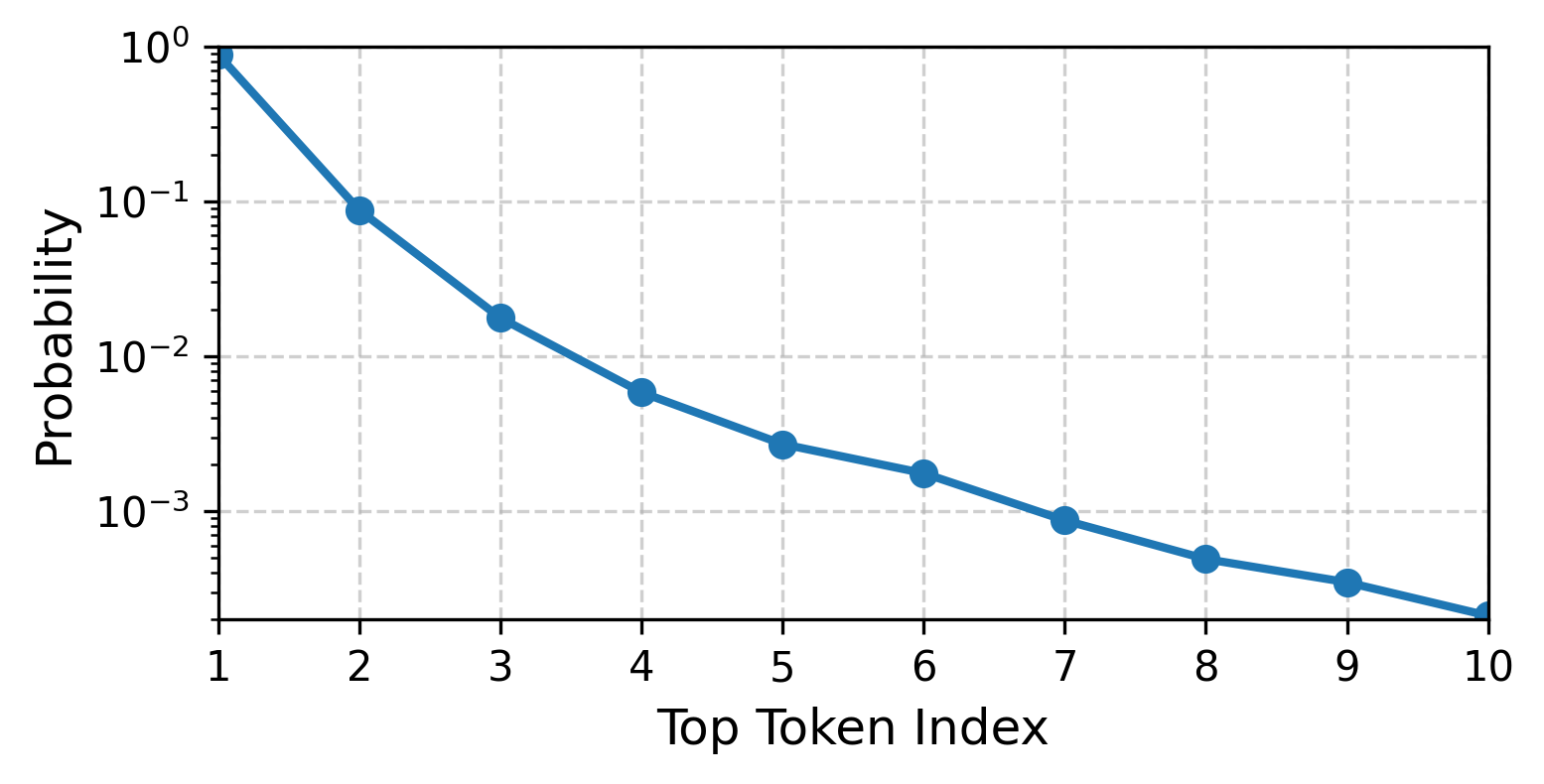}
\caption{SLM's vocabulary distribution.}
\label{fig:vocab_dist}
\end{subfigure}
\begin{subfigure}[t]{\linewidth}
\includegraphics[width=\linewidth]{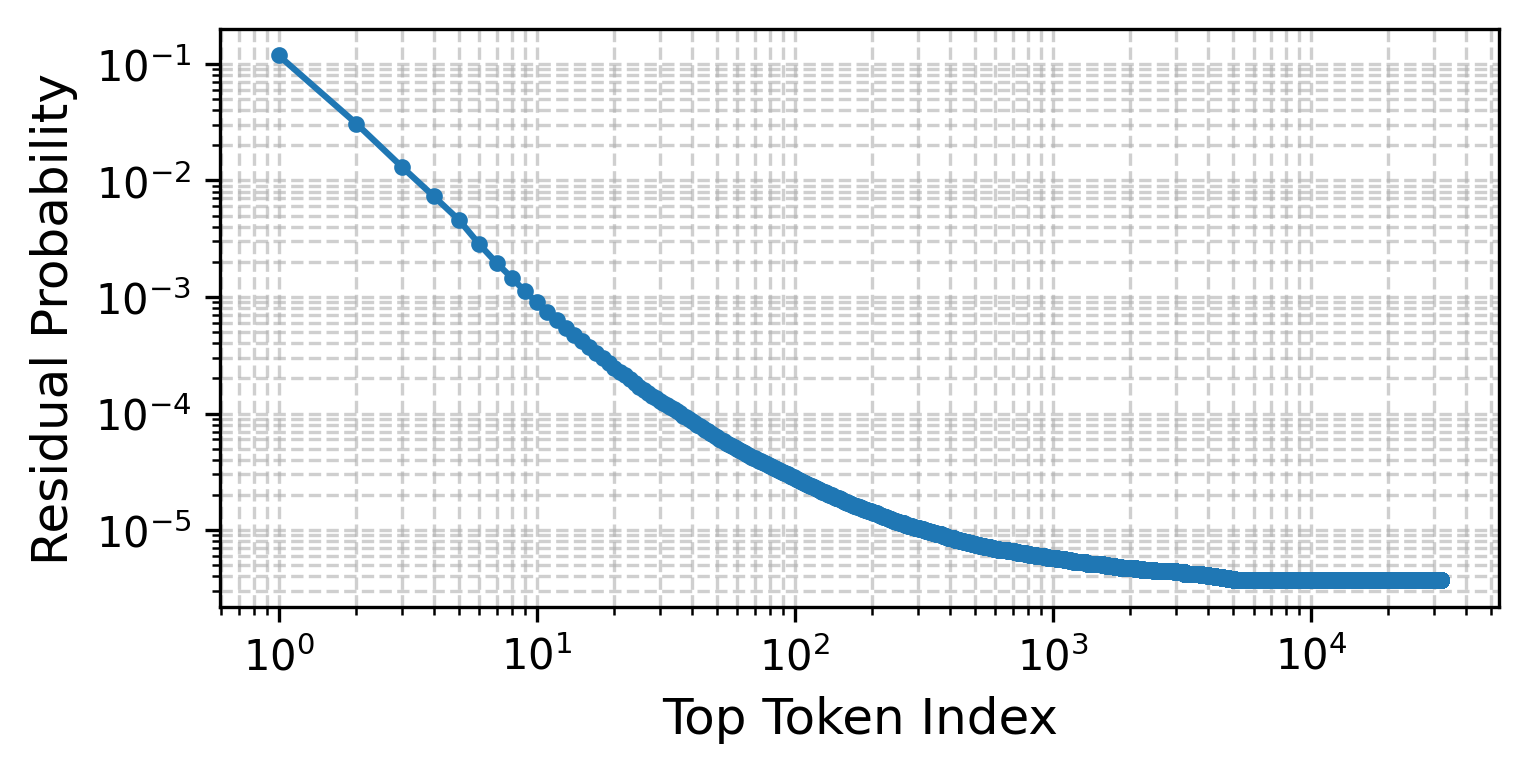}
\caption{Residual probability with respect to top token index.}
\label{fig:residual_prob}
\end{subfigure}
\caption{
(a) Average SLM vocabulary probability distribution by descending token rank (mean per rank). 
(b) Residual probability corresponding to the top token index, computed as the complementary cumulative distribution (i.e., \(1 - \sum_{i=1}^{k} x_i\)), where \(x_i\) denotes the probability of the \(k\)-th top-ranked token.
}
\label{fig:cnnvit_v}
\vspace{-10pt}
\end{figure}

To mitigate the communication bottleneck, we first examine whether the SLM’s vocabulary distribution is inherently compressible. 
\figref{fig:vocab_dist} illustrates the average token probability ranked in descending order, while \figref{fig:residual_prob} shows the residual probability—defined as the cumulative probability mass of all tokens excluding the top-ranked ones—versus the number of retained tokens.

We observe that the probability mass is highly concentrated among a small number of top-ranked tokens. 
This heavy-tailed structure suggests that truncating low-probability tokens introduces limited information loss, 
thereby motivating compressed vocabulary transmission using a top-$k$ strategy~\cite{fan2018hierarchical}.

\subsubsection{\tblue{Distortion Metric Validation}}

We now quantify the inference degradation induced by vocabulary compression.

\tblue{Recall that under the vanilla HLM design, the unbiasedness condition in~\eqref{eq:equality} 
implies $\mathsf{Bias}(t)=0$ for all $t$, according to the definition introduced in Section~III. 
This exact unbiasedness property serves as the baseline reference when introducing vocabulary compression.}

\tblue{In the absence of compression, exact unbiasedness corresponds to the primary objective:}
\begin{equation}
\tblue{\textbf{(P1)}} \quad
k(t)^*
=
\arg\min_{k(t)}
\left\{
k(t)
\;\middle|\;
\mathsf{Bias}(t)=0
\right\}.
\label{eq:primary_hlm}
\end{equation}

\tblue{
Under vocabulary compression, exact unbiasedness is generally unattainable. 
We therefore relax the accuracy requirement by allowing a bounded bias:
}
\begin{equation}
\mathsf{Bias}(t)\le\theta,
\label{eq:bias_relaxed}
\end{equation}
\begin{equation}
\tblue{\textbf{(P2)}} \quad
k(t)^*
=
\arg\min_{k(t)}
\left\{
k(t)
\;\middle|\;
\mathsf{Bias}(t)\le\theta
\right\}.
\label{eq:bias_relaxed_opt}
\end{equation}

\tblue{
Direct evaluation of $\mathsf{Bias}(t)$ remains intractable due to its dependence on the full LLM distribution and coupled rejection--resampling dynamics.
To obtain a tractable surrogate, we consider the total variation distance (TVD) between the original and distorted resampling distributions:
}
\begin{equation}
D_{\text{TV}}(\bm{p}(t),\bm{q}(t))
=
\frac{1}{2}\sum_i |p_i(t)-q_i(t)|.
\label{eq:TVD}
\end{equation}

\begin{figure}[t]
\centering
\includegraphics[width=\linewidth]{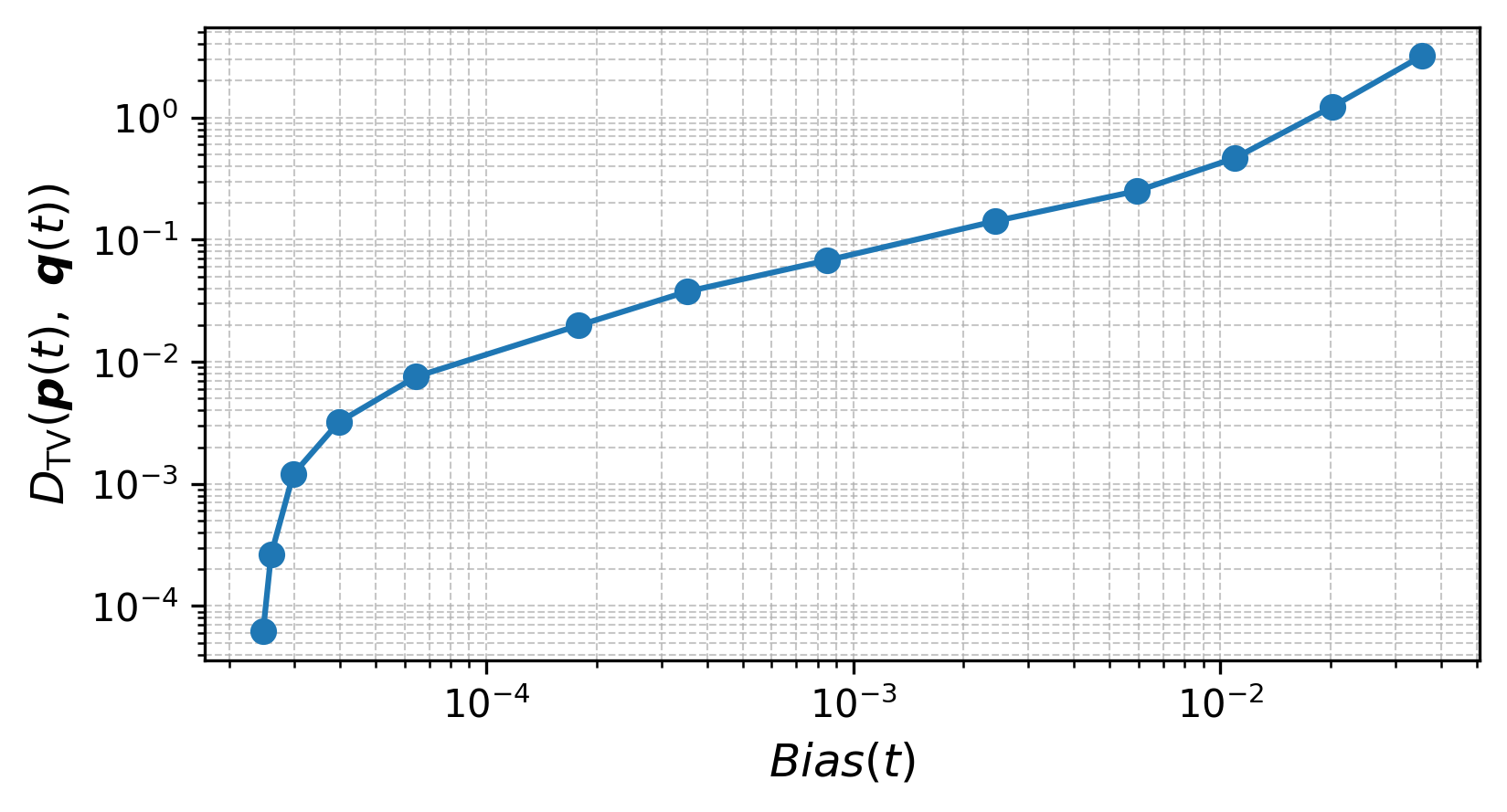}
\caption{Correlation between bias and the total variation distance (TVD) of the resampling distributions, evaluated across vocabulary sizes \( k \in [10^0, 10^4] \).}
\label{fig:topk_11}
\vspace{-10pt}
\end{figure}

Empirical analysis (Fig.~\ref{fig:topk_11}) reveals a strong linear correlation between $\mathsf{Bias}(t)$ and 
$D_{\text{TV}}(\bm{p}(t),\bm{q}(t))$ (correlation coefficient = 0.9763), 
supporting TVD as a reliable surrogate for inference degradation.
\tblue{
Accordingly, we replace the bias constraint in (P2) with a TVD-based surrogate constraint, yielding:
}
\begin{equation}
\tblue{\textbf{(P3)}} \quad
k(t)^*
=
\arg\min_{k(t)}
\left\{
k(t)
\;\middle|\;
D_{\text{TV}}(\bm{p}(t),\bm{q}(t))\le\theta
\right\}.
\label{eq:tvd_relaxed_opt}
\end{equation}

\tblue{
To implement (P3), we design two variants of CU-HLM:
(i) an \textit{offline} approach that selects a static $k^*$ using time-averaged statistical information, and
(ii) an \textit{online} approach that adapts $k(t)$ per round using only on-device information without server-side feedback.
}

\subsection{Offline Vocabulary Compression} \label{Subsection: Top-k}

Directly computing \( D_{\text{TV}}(\bm{p}(t), \bm{q}(t)) \) is computationally expensive, as it involves accessing the full LLM distribution \( y_v(t) \), computing the resampling distributions in \eqref{resampling} and \eqref{resampling_distortion}, and evaluating their element-wise difference. To improve tractability, we instead derive the following upper bound:

\begin{proposition}[Upper Bound on Total Variation Distance]
\label{prop:1}
Given that \(\bm{x}(t)\) and \(\bm{y}(t)\) denote the vocabulary distributions of the SLM and LLM at the \(t\)-th round, respectively, the total variation distance between the resampling distributions \( \bm{p}(t) \) and \( \bm{q}(t) \) is upper bounded as:
\begin{align}
D_{\text{TV}}(\bm{p}(t), \bm{q}(t)) \leq \underbrace{\frac{
\sum\limits_{i=k+1}^{|\mathcal{V}|} |x_i(t) - \hat{x}_i(t)|
}{
D_{\text{TV}}(\bm{x}(t), \bm{y}(t))
}}_{:= \mathrm{U}_{\text{TV}}(k,t)}. \label{eq:TVD_2}
\end{align}
\begin{proof}
See Appendix \ref{Proof_1}. 
\end{proof}
\end{proposition}
This upper bound offers a tractable approximation of the TVD, enabling efficient estimation of resampling distortion without requiring full computation.  
Note, however, that \( \mathrm{U}_{\text{TV}}(k,t) \) depends on the full LLM-side distribution \( \bm{y}(t) \), which is not accessible at the device side.  
To address this, we assume access to a statistical estimate of the long-term average of \( \mathrm{U}_{\text{TV}}(k,t) \), which can be practically obtained via periodic downlink feedback from the server sharing historical observations of \( \bm{y}(t) \).

\begin{remark}[Offline Vocabulary Compression]
We determine a fixed compressed vocabulary size \( k^* \) that minimizes uplink communication cost while ensuring that the resampling distortion remains within a specified tolerance \( \theta \).  
This is achieved by replacing the per-round constraint in~\eqref{eq:tvd_relaxed_opt} with a time-averaged upper bound \( \mathbb{E}_t[\mathrm{U}_{\text{TV}}(k,t)] \), leading to the following relaxed optimization:
\begin{equation}
k^* = \arg\min_{k} \left\{ k \;\middle|\; \mathbb{E}_t[\mathrm{U}_{\text{TV}}(k,t)] \leq \theta \right\}.
\end{equation}
\end{remark}
This approach is effective for offline tuning, but its applicability is limited in online contexts, as \( k^* \) is fixed across all rounds \( t \).  
Future work may explore tighter theoretical bounds using known inequalities that relate TVD to Wasserstein distance~\cite{gibbs2002choosing} or Kullback–Leibler divergence~\cite{canonne2022short,niu2025rate}.

\subsection{Online Vocabulary Compression} \label{Subsection: Top-k_v2}

To overcome the limitations of offline compression and support real-time deployment, we now propose an online vocabulary compression scheme that can be executed entirely on-device without requiring any server-side feedback.

We begin by approximating the denominator of \( \mathrm{U}_{\text{TV}}(k,t) \)—specifically, the TVD \( D_{\text{TV}}(\bm{x}(t), \bm{y}(t)) \)—as follows:
\begin{align}
D_{\text{TV}}(\bm{x}(t), \bm{y}(t)) 
&\notag \overset{\text{(a)}}{=} \sum_{i=1}^{|\mathcal{V}|} x_i(t) \left( \frac{y_i(t)}{x_i(t)} - 1 \right)^+ \\
&\overset{\text{(b)}}{\approx} \sum_{i=1}^{|\mathcal{V}|} x_i(t) \cdot \ell\left( \frac{y_i(t)}{x_i(t)}-1\right), \label{eq:approx}
\end{align}
where \( \ell(z) := \frac{\ln(1 + e^{\eta z})}{\eta} \) denotes the softplus function with temperature parameter \( \eta \). Step (a) is valid since \( x_i(t) > 0 \) from the softmax in \eqref{eq:local_process}, and (b) uses a smooth approximation of the ReLU function~\cite{nair2010rectified}.

The maximum error of this approximation is bounded by \( \ln 2 / \eta \), occurring at \( z = 0 \), i.e., when \( y_i(t) = x_i(t) \) (zero rejection). Since such cases are skipped under U-HLM, the approximation error is negligible for moderate values of \( \eta \).

We define \( \hat{\mathrm{U}}_{\text{TV}}(k,t) \) as an approximate form of \( \mathrm{U}_{\text{TV}}(k,t) \), where the denominator \( D_{\text{TV}}(\bm{x}(t), \bm{y}(t)) \) is replaced by its approximated counterpart in~\eqref{eq:approx}.
 We now present an upper bound on this term:
\begin{proposition}[Approximated Upper Bound on \( \mathrm{U}_{\text{TV}}(k,t) \)]
\label{prop:2}
The approximated TVD \( \hat{\mathrm{U}}_{\text{TV}}(k,t) \) is upper bounded as:
\begin{align}
\hat{\mathrm{U}}_{\text{TV}}(k,t) < \underbrace{\frac{
\sum\limits_{i=k+1}^{|\mathcal{V}|} |x_i(t) - \hat{x}_i(t)|
}{
(1 - x_d(t)) \cdot \ell(-1) + x_d(t) \cdot \ell(-\beta_d(t))
}}_{:= \overline{\mathrm{U}}_{\text{TV}}(\beta_d(t))}. \label{eq:prop_2}
\end{align}
\begin{proof}
See Appendix \ref{Proof_2}. 
\end{proof}
\end{proposition}

All components of \( \overline{\mathrm{U}}_{\text{TV}}(\beta_d(t)) \)—namely \( x_d(t) \), \( \beta_d(t) \), and \( \sum_{i=k+1}^{|\mathcal{V}|} |x_i(t) - \hat{x}_i(t)| \)—are either directly observable or locally computable on the device. In particular, \( \beta_d(t) \) is estimated from the token's uncertainty using the linear model in~\eqref{eq:uncertainty_rejection2}. This enables fully on-device, token-level vocabulary compression without requiring server-side information.

\begin{remark}[Uncertainty-Aware Online Vocabulary Compression]
We propose an online vocabulary compression scheme that selects the compressed vocabulary size at each round \( t \) as:
\begin{equation}
k(t)^* 
= \arg\min_{k(t)} \left\{
k(t) \;\middle|\; 
\overline{\mathrm{U}}_{\text{TV}}(au(t)+b) \leq \theta
\right\}. \label{eq:online_policy}
\end{equation}
The goal is to minimize the instantaneous uplink payload while ensuring that the estimated resampling distortion remains within the tolerance \( \theta \).  
This formulation serves as a relaxed version of the original problem in~\eqref{eq:tvd_relaxed_opt}, where the intractable constraint on \( D_{\text{TV}} \) is approximated using observable, uncertainty-based quantities.
\end{remark}
\noindent Since \( \overline{\mathrm{U}}_{\text{TV}}(\beta_d(t)) \) increases with \( \beta_d(t) \), the resulting vocabulary size \( k(t)^* \) naturally grows with uncertainty, thereby allocating higher transmission fidelity to more uncertain tokens.


\begin{figure}[t]
\centering
\includegraphics[width=\linewidth]{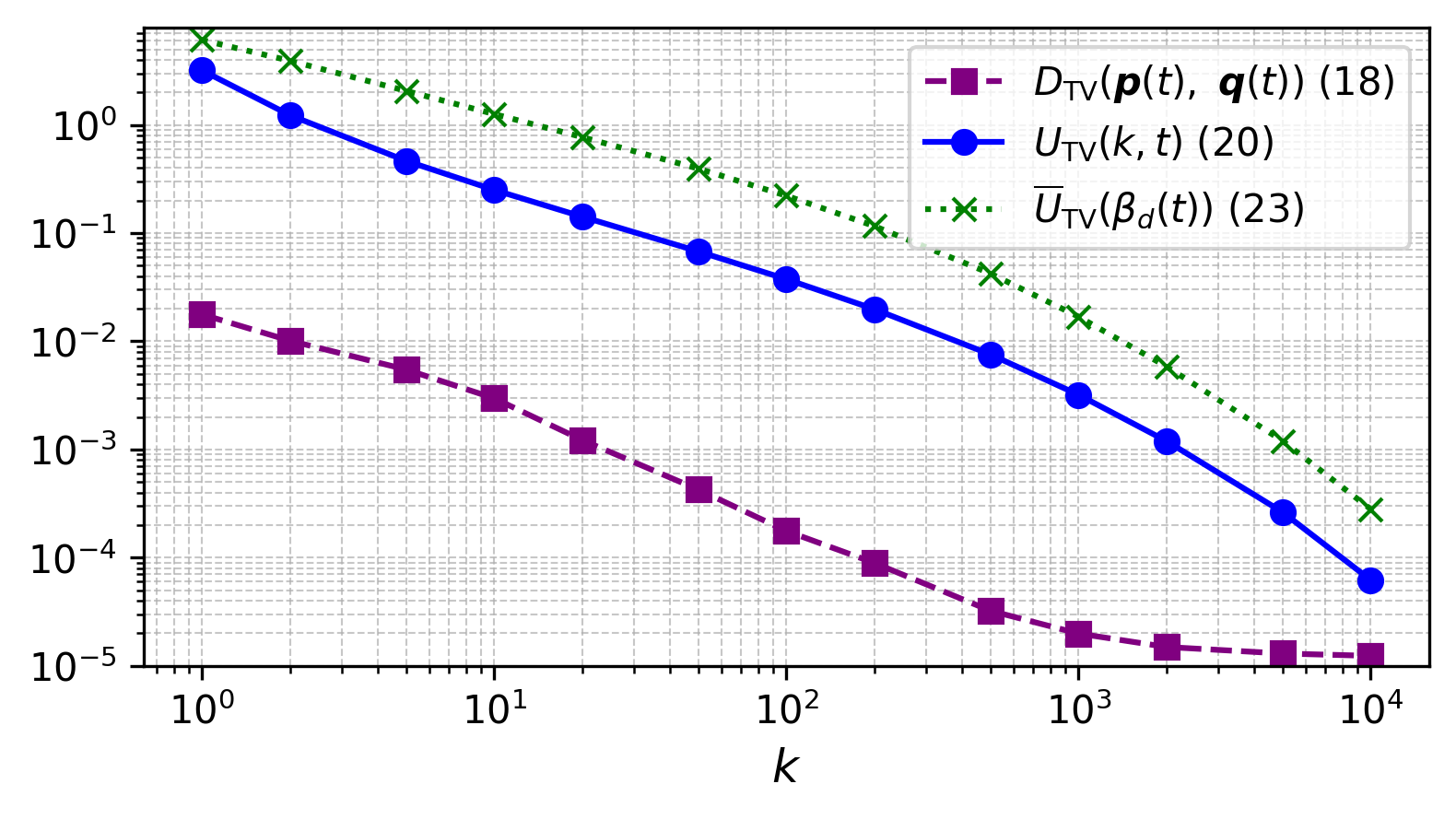}
\caption{Comparison of TVD and its upper bounds w.r.t. $k$.}
\label{fig:topk_21}
\vspace{-10pt}
\end{figure}

We note that both bounds \( \mathrm{U}_{\text{TV}}(k,t) \) and \( \overline{\mathrm{U}}_{\text{TV}}(\beta_d(t)) \) share the same numerator—dependent on the residual mass beyond the top-\( k \) tokens—and differ only in their denominator formulations.  
As a result, both bounds exhibit similar decreasing trends as the compressed vocabulary size \( k \) increases.  
As shown in \figref{fig:topk_21}, \( \mathrm{U}_{\text{TV}}(k,t) \) provides a tight upper bound on the actual TVD, while \( \overline{\mathrm{U}}_{\text{TV}}(\beta_d(t)) \) also becomes increasingly accurate with larger vocabulary sizes, despite relying only on local information.  

While both the offline and online designs aim to minimize communication under a fixed distortion constraint, our framework can be extended to the dual objective—maximizing inference accuracy under a communication budget—which we leave for future work. 

Although our analysis is based on top-ranked token selection, the proposed vocabulary size adaptation strategy can also be extended to threshold-based methods such as top-\(p\) (nucleus) sampling~\cite{holtzman2019curious} and min-\(p\) sampling~\cite{nguyen2024turning}, using the conversion rules outlined in~\cite{tang2024top}.

\tblue{Furthermore, while our framework is developed on top of the HLM-style verification mechanism, the proposed opportunistic skipping and compression principles are not inherently tied to this specific architecture. For example, methods such as EAGLE \cite{li2025eagle} exhibit algorithm-level opportunistic decoding behavior, where draft tokens with higher acceptance likelihood are prioritized. This structural property makes such architectures naturally compatible with opportunistic skipping and compression strategies. A detailed integration and systematic evaluation of these extensions are left for future work.}




These results lay the groundwork for the offline and online vocabulary compression strategies in CU-HLM.
Through tractable upper bound formulations and corresponding policy designs, we are equipped to empirically evaluate their effectiveness in the subsequent experiments.



\vspace{-10pt}
\section{Numerical Evaluation}\label{sec:eval}

This section presents experimental results validating the effectiveness of the proposed CU-HLM framework, including both its online and offline vocabulary compression schemes. We begin by detailing the simulation setup and evaluation metrics, followed by a comprehensive analysis of the results.

\vspace{-10pt}
\subsection{Simulation Setup}

Unless otherwise specified, all experiments are conducted using TinyLlama-1.1B as the SLM and Llama2-13B as the LLM, with a vocabulary size of \( |\mathcal{V}| = 32{,}000 \). The Alpaca dataset~\cite{alpaca} is employed as the benchmark, with a fixed set of 100 randomly selected samples used across all experiments. All simulations are performed on a Linux-based server equipped with an 8-core Intel Xeon Silver 4215R CPU and three Nvidia GeForce RTX 3090 GPUs. Under this setup, the average per-token wall time is measured as \( \tau_{\text{SLM}} = 25.6\,\text{ms} \) and \( \tau_{\text{LLM}} = 104.6\,\text{ms} \).

We evaluate the proposed CU-HLM framework—both with online and offline vocabulary compression—alongside U-HLM, and compare them against several baselines: (i) LLM-only inference, (ii) SLM-only inference, (iii) the original HLM, (iv) Rand-HLM, which skips uplink transmissions at random with a fixed probability of 0.5, \tblue{and (v) TK-SLT~\cite{zheng2025communication}, which performs vocabulary compression with a fixed truncated vocabulary size of $320$.} The performance is evaluated using the following metrics:
\begin{figure*}
\begin{minipage}{0.51\textwidth}
    \centering
    \includegraphics[width=\textwidth]{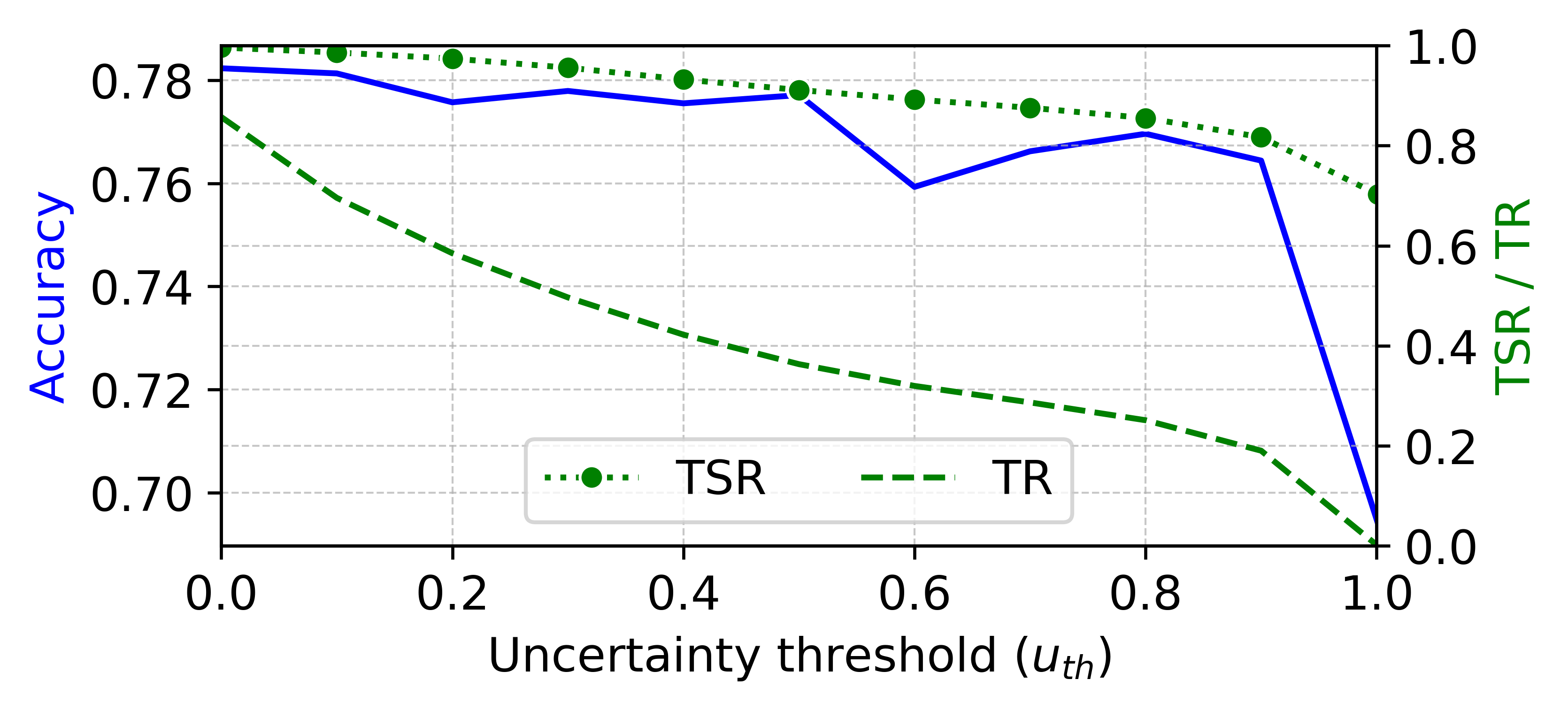}
    \subcaption{Accuracy, transmission rate (TR), and true skip rate (TSR) as functions of the uncertainty threshold $u_{\text{th}}$.}
    \label{fig:u_3}
\end{minipage}
\hfill
\begin{minipage}{0.48\textwidth}
    \centering
    \includegraphics[width=\textwidth]{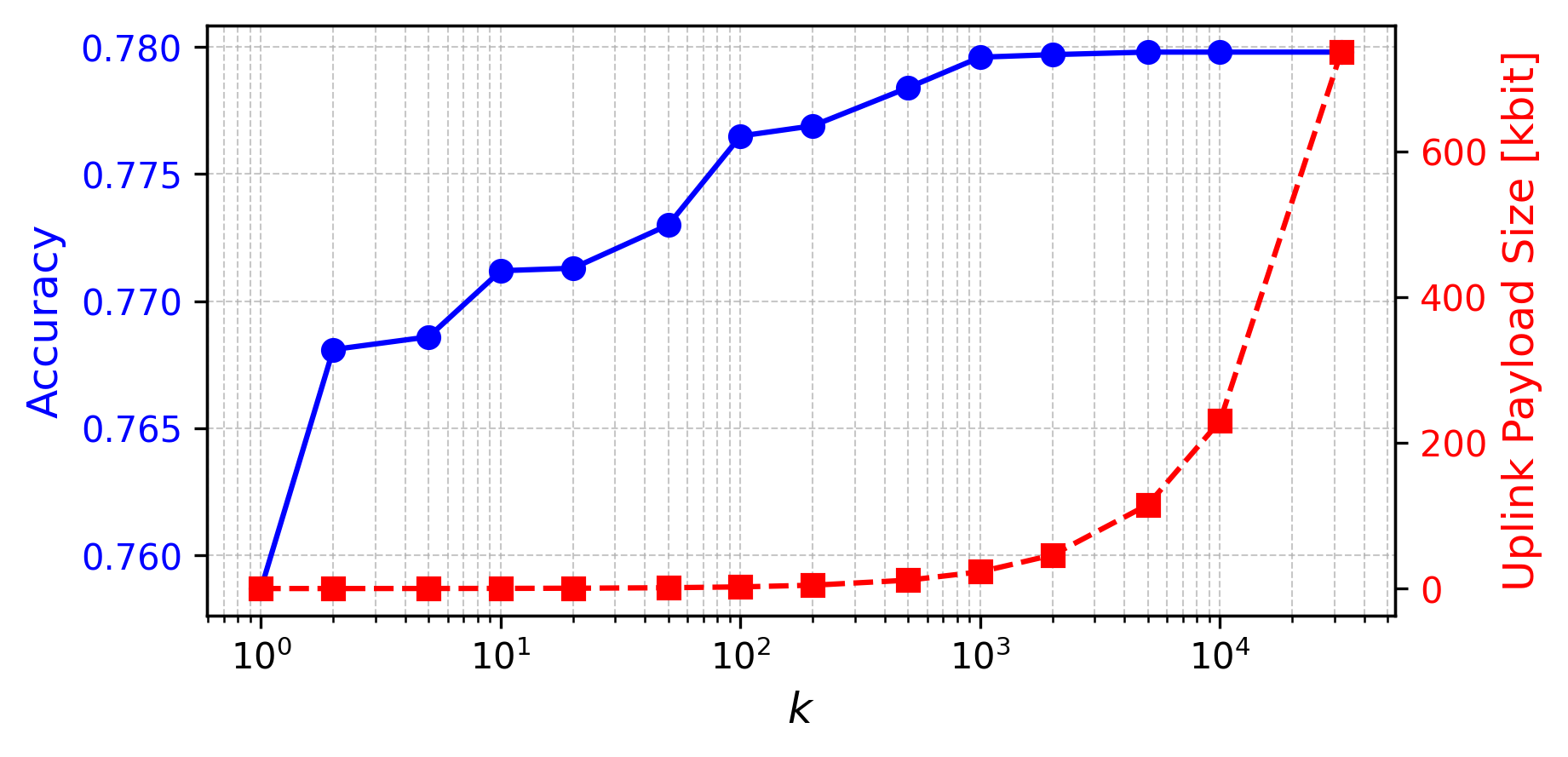}
    \subcaption{Accuracy and uplink payload size versus compressed vocabulary size $k$.}
    \label{fig:topk_3}
\end{minipage}
\caption{Accuracy and communication efficiency with respect to design parameters in CU-HLM.}
\label{fig:combined2}
\vspace{-10pt}
\end{figure*}

\begin{itemize}
    \item \textbf{Inference Accuracy:} Measured as the cosine similarity between sentence embeddings of the generated response and the corresponding ground-truth answer, where embeddings are computed using a BERT model~\cite{alaparthi2020bidirectional}.
    
    \item \textbf{Token Throughput:} Defined as the average number of tokens generated per second over the full sequence, extended from the per-round throughput defined in \eqref{eq:spec_latency}.

\end{itemize}

The remaining hyperparameters are configured as: uplink bandwidth \( W = 10\,\text{MHz} \), \( b_{\text{prob}} = 8 \), and \( r_{\text{max}} = 512 \).

\subsection{Impact of Design Parameters on Accuracy and communication efficiency}

Before evaluating the overall performance of CU-HLM, this subsection investigates the empirical impact of its two key hyperparameters—namely, the uncertainty threshold and the compressed vocabulary size—on inference accuracy and communication efficiency. Based on this analysis, we finalize the parameter settings for subsequent experiments.

To begin, we assess the effect of the uncertainty threshold \( u_{\text{th}} \) using the following two metrics:
\begin{itemize}
    \item \textbf{Transmission Rate (TR):} The proportion of rounds in which an uplink transmission occurs. In standard HLM, this defaults to 1 and serves as an indicator of communication efficiency.
    \item \textbf{True Skip Rate (TSR):} The probability that a skipped token is eventually accepted by the LLM, used to assess the accuracy impact of skipping.
\end{itemize}

\figref{fig:u_3} shows that increasing \( u_{\text{th}} \) leads to a decrease in both TR and TSR. A higher threshold makes the model more aggressive in skipping, thereby reducing the frequency of uplink transmissions (lower TR). However, this also increases the likelihood of skipping tokens that would have been rejected (lower TSR). Fortunately, the drop in TSR is relatively moderate—remaining above 0.702—even at high thresholds. This resilience can be attributed to the distribution of uncertainty values, which is concentrated near the extremes (0 and 1), as shown in \figref{fig:u_11}. This implies that raising \( u_{\text{th}} \) still preserves a significant portion of accurate skips.

As a result, inference accuracy also tends to decrease with increasing \( u_{\text{th}} \), but remains well-preserved up to a certain point. Notably, two inflection points in the accuracy curve are observed, with the most significant drop occurring at \( u_{\text{th}} = 0.9 \). This aligns well with the risk-prone threshold of 0.8117 derived in \textbf{Theorem}~\ref{theorem:combined}, thereby validating the theoretical analysis. For the uncertainty threshold, we set \( u_{\text{th}} = 0.8 \), which allows the model to skip uplink transmissions for approximately 74.8\% of tokens.

\figref{fig:topk_3} illustrates the trade-off between inference accuracy and uplink payload size with respect to the compressed vocabulary size \( k \). As expected, increasing \( k \) improves accuracy while linearly increasing communication cost, showing a clear monotonic trend in both metrics. Based on these empirical observations, and considering the trade-off between communication efficiency and accuracy, we fix the tolerance parameter as \( \theta = 0.1 \), which yields an offline vocabulary size of \( k^* = 30 \). This corresponds to less than 0.1\% of the full vocabulary size in terms of uplink payload. 

It is worth noting that the reported uplink payload size represents a worst-case estimate, indicating room for further optimization in both communication and computation. On the communication side, advanced compression techniques—such as source coding based on token-wise entropy within the vocabulary—could further reduce the transmission cost. On the computation side, our analysis does not yet account for potential runtime savings that may result from operating on smaller values of \( k \), which could further improve overall system efficiency.

\subsection{Inference Accuracy and Token Throughput Comparison}
\label{exp_perfor}

\begin{figure*}[t]
\centering
\begin{minipage}{0.49\textwidth}
    \centering
    \includegraphics[width=\textwidth]{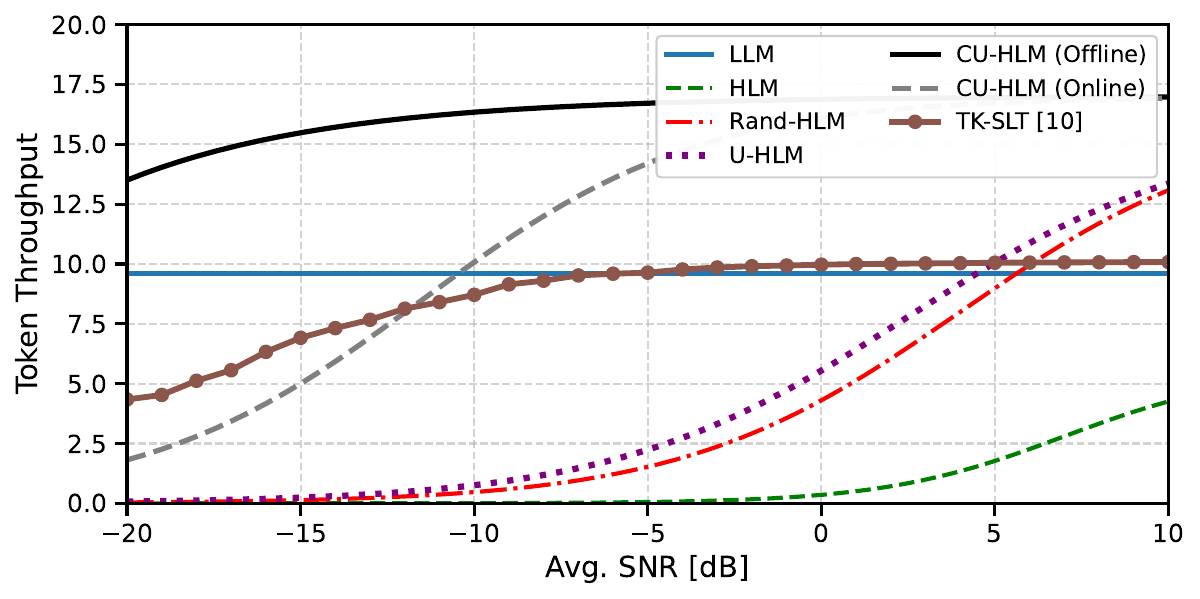}
    \subcaption{\tblue{Rayleigh fading channel.}}
    \label{fig:rayleigh}
\end{minipage}
\hfill
\begin{minipage}{0.49\textwidth}
    \centering
    \includegraphics[width=\textwidth]{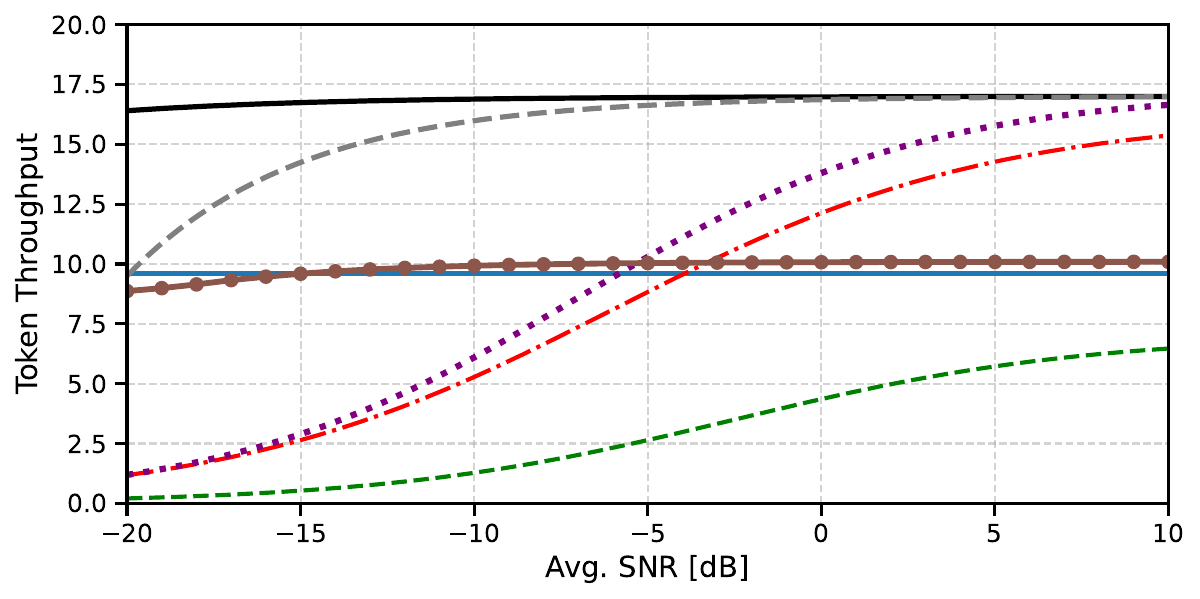}
    \subcaption{\tblue{Rician fading channel with a K-factor of $10$\,dB.}}
    \label{fig:rician}
\end{minipage}
\caption{Token throughput across inference methods as a function of average SNR under different fading conditions.}
\label{fig:combined}
\vspace{-10pt}
\end{figure*}

This subsection presents the overall performance of CU-HLM by jointly assessing the impact of previously discussed components. Table~\ref{table:inference_accuracy_llm} evaluates inference accuracy across multiple model configurations and datasets—including QED, CREAK, and StrategyQA from the FLAN collection~\cite{longpre2023flan}, \tblue{as well as Databricks Dolly 15K—while extending the evaluation to include Llama2-7B as the LLM.} \figref{fig:combined} further illustrates token throughput across different inference methods under two wireless fading environments: Rayleigh and Rician (with $K = 10$ dB), across average SNRs ranging from $-20$ to $10$ dB. SLM is omitted from the token throughput curves in \figref{fig:combined}, 
since it operates without wireless transmission and thus remains unaffected by channel conditions. 
Its throughput is $39$ tokens/s, determined solely by its standalone computation time, 
thereby serving as a computation-only upper bound in token throughput.

We first note that HLM and LLM serve as upper bounds in inference accuracy, 
as observed in Table~\ref{table:inference_accuracy_llm}. In general, better channel conditions—i.e., higher SNR and under a Rician fading channel—lead to improved token throughput, as demonstrated in \figref{fig:rayleigh} and \figref{fig:rician}. \tblue{Within this context, U-HLM outperforms Rand-HLM by consistently achieving higher inference accuracy at similar token throughput levels, while CU-HLM (Offline) achieves substantially higher token throughput than TK-SLT with comparable inference accuracy, demonstrating the efficacy of uncertainty-based opportunistic transmission on inference accuracy and token throughput, respectively.}

\begin{table*}[t!]
    \centering
    \caption{Comparison of inference accuracy across different methods, SLM--LLM pairs, and datasets.}
    \resizebox{0.9\textwidth}{!}{%
    \begin{tabular}{lcccccccccc}
        \toprule
        \multirow{3}{*}{Inference Method} 
        & \multicolumn{5}{c}{\textbf{TinyLlama 1.1B -- Llama2 13B}} & \multicolumn{5}{c}{\textbf{TinyLlama 1.1B -- Llama2 7B}} \\
        \cmidrule(lr){2-6} \cmidrule(lr){7-11}
        & \textbf{Alpaca} & \textbf{QED} & \textbf{CREAK} & \textbf{StrategyQA} & \tblue{\textbf{Dolly 15K}}
        & \textbf{Alpaca} & \textbf{QED} & \textbf{CREAK} & \textbf{StrategyQA} & \tblue{\textbf{Dolly 15K}} \\
        \midrule
        LLM & 0.7782 & 0.6954 & 0.6722 & 0.7122 & \tblue{0.5801} & 0.7758 & 0.6271 & 0.6621 & 0.6779 & \tblue{0.5835}\\
        SLM & 0.6941 & 0.5139 & 0.5291 & 0.5694 & \tblue{0.4513} & 0.6521 & 0.5139 & 0.5291 & 0.5694 & \tblue{0.4513}\\
        HLM & 0.7825 & 0.7226 & 0.6946 & 0.6983 & \tblue{0.5747} & 0.7823 & 0.6517 & 0.6529 & 0.6883 & \tblue{0.5768}\\
        \midrule
        Rand-HLM & 0.7385 & 0.6128 & 0.6406 & 0.6227 & \tblue{0.5327} & 0.7382 & 0.5534 & 0.5682 & 0.6200 & \tblue{0.5399}\\
        \tblue{TK-SLT \cite{zheng2025communication}} & \tblue{0.7692} &\tblue{0.6798} & \tblue{0.6499} & \tblue{0.6818} & \tblue{0.5533} &\tblue{0.7302} & \tblue{0.6304}&\tblue{0.6308} &\tblue{0.6522} & \tblue{0.5598}\\
        U-HLM & 0.7704 & 0.7122 & 0.6854 & 0.6844 & \tblue{0.5554} & 0.7696 & 0.6381 & 0.6550 & 0.6532 & \tblue{0.5703} \\
        \textbf{CU-HLM (Offline)} & 0.7583 & 0.6713 & 0.6508 & 0.6773 & \tblue{0.5519} & 0.7455 & 0.6130 & 0.6062 & 0.6242 & \tblue{0.5594} \\
        \textbf{CU-HLM (Online)} & 0.7623 & 0.6876 & 0.6672 & 0.6817 & \tblue{0.5525} & 0.7622 & 0.6387 & 0.6460 & 0.6409 & \tblue{0.5672} \\
        \bottomrule
    \end{tabular}
    }
    \label{table:inference_accuracy_llm}
    \vspace{-10pt}
\end{table*}

Next, we analyze CU-HLM with offline and online vocabulary compression. Compared to standard HLM inference, CU-HLM significantly improves token throughput while maintaining competitive accuracy. For instance, on the Alpaca dataset with Llama2-13B, CU-HLM achieves inference accuracy of 96.91\% (offline) and 97.42\% (online). The token throughput gains are consistently observed across various fading channels and SNRs.  
As shown in \figref{fig:rayleigh}, the improvement is particularly significant under Rayleigh fading with an average SNR of $-20\,\text{dB}$, where CU-HLM achieves up to $1014.1\times$ (offline) and $206\times$ (online) throughput gains.  
These gains are primarily attributed to the uncertainty-aware opportunistic and compressed transmission, which substantially reduces both uplink transmit opportunity and payload size.

The accuracy--throughput trade-off between the offline and online variants of CU-HLM arises from differences in the tightness of their respective compression bounds. CU-HLM (Offline) employs a fixed compressed vocabulary size of \( k^* = 30 \), whereas CU-HLM (Online) dynamically adjusts \( k(t)^* \) between 4 and 5{,}715 (averaging 832, i.e., only 2.6\% of the full vocabulary) based on the estimated rejection probability and $x_d(t)$, as defined in~\eqref{eq:prop_2}. While the bound used in CU-HLM (Online) is looser, its ability to dynamically select near-optimal compression levels allows it to achieve token throughput comparable to its offline counterpart under moderate SNR conditions.  
However, under extremely low-SNR regimes, CU-HLM (Offline), which leverages server-side information, yields more stable gains, though the gap diminishes as the SNR increases and channel quality improves.

Taken together, the results in Table~\ref{table:inference_accuracy_llm} and \figref{fig:combined} demonstrate that CU-HLM offers a favorable balance between inference accuracy and communication efficiency. It generalizes well across diverse datasets, model configurations, and wireless channel conditions. The online variant operates solely on on-device information without statistical feedback, yet achieves performance comparable to the offline variant, which leverages tighter, feedback-assisted compression. \tblue{Additional token-level similarity metrics and outage-channel evaluations are provided in Appendices~\ref{append:c} and \ref{append:d}, respectively, showing consistent trends with the main results.}

\begin{table}[t!]
    \centering
    \caption{Latency breakdown and token throughput of U-HLM and CU-HLM (Online) under different configurations, measured under Rayleigh fading with average SNR = 10\,dB.  
    SLM computation time is fixed at 25.6\,ms (64\,ms for 7B--13B*), and all values reflect average latency per token.}
    \begin{tabular}{lccc}
        \toprule
        \textbf{Method} & \textbf{Communication} & \makecell{\textbf{LLM} \\ \textbf{Computation}} & \makecell{\textbf{Token} \\ \textbf{Throughput}} \\
        \midrule
        \multicolumn{4}{l}{\textit{Baseline: U-HLM}} \\
        \quad + No KD & 6.3\,ms & 31.1\,ms & 15.9 \\
        \quad + KD & 5.9\,ms & 29.0\,ms & 16.5 \\
        \quad + 7B--13B* & 2.8\,ms & 13.8\,ms & 12.4 \\
        \midrule
        \multicolumn{4}{l}{\textit{CU-HLM (Online)}} \\
        \quad + No KD & 38.1\,$\mu$s & 30.0\,ms & 18.0 \\
        \quad + KD & 35.5\,$\mu$s & 27.9\,ms & \textbf{18.7} \\
        \quad + 7B--13B* & 16.2\,$\mu$s & 12.8\,ms & 13.0 \\
        \bottomrule
    \end{tabular}
    \label{table:latency_breakdown_smart}
    \vspace{-10pt}
\end{table}

\subsection{Ablation Study: Impact of SLM Architecture and Alignment}
\label{subsec:acceptance_future}

The token throughput of CU-HLM is fundamentally influenced by how well the SLM's predictions align with those of the LLM. While CU-HLM significantly reduces uplink communication latency through uncertainty-aware skipping and compression, the LLM computation latency remains a dominant bottleneck, as shown in Table~\ref{table:latency_breakdown_smart}. To examine whether improved SLM behavior can further alleviate this issue, we evaluate two variations designed to enhance either the SLM architecture or its alignment with the LLM. For architectural improvement, we replace the default SLM with a larger 7B model. For alignment, we fine-tune the SLM using knowledge distillation (KD), where the LLM (Llama2-13B) provides soft targets in the form of output logits.
\tblue{We adopt Kullback–Leibler (KL) divergence as the distillation loss, with a distillation temperature $\tau = 1$ and a distillation ratio $\lambda = 0.1$. The student model is fine-tuned using LoRA ($r=12$, $\alpha=32$, dropout $0.1$), while keeping the teacher model frozen.}

Experiments under Rayleigh fading with 10\,dB SNR show that both strategies raise acceptance rates and reduce LLM invocation. KD yields a modest 2\% increase in acceptance, translating to a 7\% reduction in LLM latency without increasing SLM computation time. The larger SLM achieves a 16.5\% acceptance gain but incurs over 150\% higher SLM computation latency, offsetting part of the benefit. These results confirm that both architectural capacity and alignment quality contribute to acceptance rates, and thus token throughput, with distinct trade-offs. Among the two, KD provides a lightweight yet effective improvement, and may be further extended to training-time integration~\cite{peng2024pre} or applied alongside sparse transfer techniques~\cite{zaidi2025sparse}. We leave this direction as promising future work.

\section{Conclusion} \label{sec:con}
This paper presents CU-HLM, a communication-efficient hybrid language model framework tailored for wireless edge inference. By leveraging the strong empirical correlation between token-level uncertainty and rejection probability, we develop an uncertainty-aware opportunistic transmission mechanism that selectively skips both uplink communication and LLM computation for low-uncertainty tokens. To further reduce communication overhead under high uncertainty, we introduce a compressed vocabulary transmission scheme and analytically formulate the optimal compression policy for both offline and online deployment scenarios. Extensive experiments across diverse datasets, model configurations, and wireless channel conditions confirm that CU-HLM substantially enhances communication efficiency while maintaining near-LLM inference accuracy. These results highlight CU-HLM's potential as a scalable solution for efficient on-device language modeling in bandwidth-constrained environments. Future work includes extending our approach to heterogeneous multi-agent systems built upon A2A and MCP \cite{radosevich2025mcp} protocols, where agents differ in capability, context access, and model capacity.


\vspace{-10pt}
\appendices
\section{Proof of Proposition~\ref{prop:1}}\label{Proof_1}

We begin by recalling the definition of the TVD between two probability distributions \( \bm{p}(t) \) and \( \bm{q}(t) \) as
\begin{equation}
    D_{\text{TV}}(\bm{p}(t), \bm{q}(t)) = \frac{1}{2} \left\| \bm{p}(t) - \bm{q}(t) \right\|_1, \label{eq:proof_1}
\end{equation}
where \( \|\cdot\|_1 \) denotes the \( \ell_1 \)-norm, defined as the sum of the absolute values of the vector’s components.

To simplify the analysis, define two auxiliary vectors \( \bm{\gamma} \) and \( \bm{\omega} \) such that the \(v\)-th components are given by \( \gamma_v = (y_v(t) - x_v(t))^+ \) and \( \omega_v = (y_v(t) - \hat{x}_v(t))^+ \), respectively.

Substituting these into \eqref{eq:proof_1}, we have:
\begin{equation}
    D_{\text{TV}}(\bm{p}(t), \bm{q}(t)) = \frac{1}{2} \left\| \frac{\bm{\gamma}}{\|\bm{\gamma}\|_1} - \frac{\bm{\omega}}{\|\bm{\omega}\|_1} \right\|_1.
\end{equation}

We decompose the \( \ell_1 \)-norm as follows:
{\footnotesize
\begin{align}
 \left\|  \frac{\bm{\gamma}}{\|\bm{\gamma}\|_1} - \frac{\bm{\omega}}{\|\bm{\omega}\|_1} \right\|_1 
 &= \left\| \left(  \frac{\bm{\gamma}}{\|\bm{\gamma}\|_1} - \frac{\bm{\omega}}{\|\bm{\gamma}\|_1} \right) 
 + \left(  \frac{\bm{\omega}}{\|\bm{\gamma}\|_1} - \frac{\bm{\omega}}{\|\bm{\omega}\|_1} \right) \right\|_1 \\
 &\overset{\text{(a)}}{\leq} \left\|  \frac{\bm{\gamma} - \bm{\omega}}{\|\bm{\gamma}\|_1} \right\|_1 
 + \left\| \bm{\omega} \left( \frac{1}{\|\bm{\gamma}\|_1} - \frac{1}{\|\bm{\omega}\|_1} \right) \right\|_1,
\end{align}}
where (a) follows from the triangle inequality.

We now bound each term separately. For the first term:
\begin{equation}\label{append_first}
\left\| \frac{\bm{\gamma} - \bm{\omega}}{\|\bm{\gamma}\|_1} \right\|_1 
= \frac{\|\bm{\gamma} - \bm{\omega}\|_1}{\|\bm{\gamma}\|_1}.
\end{equation}

For the second term, using the reverse triangle inequality:
\begin{align}\label{append_second}
\left\| \bm{\omega} \left( \frac{1}{\|\bm{\gamma}\|_1} - \frac{1}{\|\bm{\omega}\|_1} \right) \right\|_1 
&= \|\bm{\omega}\|_1 \cdot \left| \frac{1}{\|\bm{\gamma}\|_1} - \frac{1}{\|\bm{\omega}\|_1} \right|\nonumber\\
&= \frac{|\|\bm{\gamma}\|_1 - \|\bm{\omega}\|_1|}{\|\bm{\gamma}\|_1} \nonumber\\
&\leq \frac{\|\bm{\gamma} - \bm{\omega}\|_1}{\|\bm{\gamma}\|_1}.
\end{align}

Thus, combining \eqref{append_first} and \eqref{append_second}, we obtain:
\begin{equation}
D_{\text{TV}}(\bm{p}(t), \bm{q}(t)) 
\leq \frac{\|\bm{\gamma} - \bm{\omega}\|_1}{\|\bm{\gamma}\|_1}.
\end{equation}

Next, we bound the numerator. Since ReLU is a 1-Lipschitz function, it holds that:
\begin{align}
  \|\bm{\gamma} - \bm{\omega}\|_1 &\leq \|\bm{x}(t) - \hat{\bm{x}}(t)\|_1 \nonumber\\
  &= \left\| \left(\bm{x}(t) - \hat{\bm{x}}(t) \right)_{k+1:|\mathcal{V}|} \right\|_1,  \label{append_nume}
\end{align}
where \( (\cdot)_{a:b} \) denotes the subvector consisting of indices from \( a \) to \( b \), exploiting the fact that \( \bm{x}(t) \) and \( \hat{\bm{x}}(t) \) are identical over the top-\( k \) elements.

Finally, for the denominator, since both \( \bm{x}(t) \) and \( \bm{y}(t) \) are valid probability distributions, we apply mass conservation to obtain:
\begin{align}
\|\bm{\gamma}\|_1&=\sum_{i=1}^{|\mathcal{V}|} (y_i(t) - x_i(t))^+\\
&= \frac{1}{2} \sum_{i=1}^{|\mathcal{V}|} |y_i(t) - x_i(t)| = D_{\text{TV}}(\bm{x}(t), \bm{y}(t)). \label{append_denom}
\end{align}
Substituting \eqref{append_nume} and \eqref{append_denom} completes the proof. \qed



\vspace{-10pt}
\section{Proof of Proposition~\ref{prop:2}} \label{Proof_2}

From the approximation in~\eqref{eq:approx}, we decompose the sum into the draft token and the non-draft tokens as:
\begin{align}
   \sum_{i=1}^{|\mathcal{V}|} x_i(t) \cdot \ell\left( \frac{y_i(t)}{x_i(t)} - 1 \right)
= x_d(t) \cdot \ell\left( \frac{y_d(t)}{x_d(t)} - 1 \right) \notag
\\ + \sum_{i \neq d} x_i(t) \cdot \ell\left( \frac{y_i(t)}{x_i(t)} - 1 \right). 
\end{align}

Since \( \ell(z) \) is monotonically increasing and both \( x_i(t) \) and \( y_i(t) \) are outputs of softmax operations (thus strictly positive for all \( i \in \mathcal{V} \)), we separately bound the draft and non-draft terms as follows.

For the draft token term, we have:
\begin{align}
\ell\left( \frac{y_d(t)}{x_d(t)} - 1 \right)
&= \ell\left( -\left(1-\frac{y_d(t)}{x_d(t)}\right) \right) \notag \\
&\geq \ell\left( -\left(1-\frac{y_d(t)}{x_d(t)}\right)^+ \right) \nonumber\\
&= \ell\left(-\beta_d(t)\right),
\end{align}
where the inequality follows because \(\ell(z)\) is increasing and \( \left(1-\frac{y_d(t)}{x_d(t)}\right) \leq \left(1-\frac{y_d(t)}{x_d(t)}\right)^+ \). Next, for each non-draft token \( i \neq d \), since \( y_i(t) > 0 \) and \( x_i(t) > 0 \), we have:
\begin{equation}
\ell\left( \frac{y_i(t)}{x_i(t)} - 1 \right) > \ell(-1).
\end{equation}

Combining the two bounds, the denominator in Proposition~\ref{prop:2} is lower-bounded by
\[
x_d(t) \cdot \ell(-\beta_d(t)) + (1 - x_d(t)) \cdot \ell(-1).
\]
Substituting this into the denominator of \( \hat{\mathrm{U}}_{\text{TV}}(k,t) \) yields the desired upper bound, completing the proof. \qed








\begin{table}[t!]
    \centering
    \caption{\tblue{Token-level F1 and Jaccard similarity on Dolly 15K for TinyLlama 1.1B paired with Llama2 7B and 13B.}}
    \resizebox{1\columnwidth}{!}{%
    \begin{tabular}{lcccc}
        \toprule
        \multirow{2}{*}{\tblue{Inference Method}}
        & \multicolumn{2}{c}{\tblue{\textbf{TinyLlama 1.1B -- Llama2 7B}}}
        & \multicolumn{2}{c}{\tblue{\textbf{TinyLlama 1.1B -- Llama2 13B}}} \\
        \cmidrule(lr){2-3} \cmidrule(lr){4-5}
        & \tblue{\textbf{Token-F1}} & \tblue{\textbf{Jaccard}}
        & \tblue{\textbf{Token-F1}} & \tblue{\textbf{Jaccard}} \\
        \midrule
        \tblue{LLM}        & \tblue{\textbf{0.0555}} & \tblue{\textbf{0.0283}} & \tblue{\textbf{0.0571}} & \tblue{0.0310} \\
        \tblue{SLM}        & \tblue{0.0432} & \tblue{0.0193} & \tblue{0.0432} & \tblue{0.0193} \\
        \tblue{HLM}        & \tblue{0.0541} & \tblue{0.0280} & \tblue{0.0567} & \tblue{\textbf{0.0313}} \\
        \midrule
        \tblue{Rand-HLM}   & \tblue{0.0375} & \tblue{0.0180} & \tblue{0.0530} & \tblue{0.0243} \\
        \tblue{TK-SLT \cite{zheng2025communication}}      
                           & \tblue{0.0500} & \tblue{0.0265} & \tblue{0.0532} & \tblue{0.0286} \\
        \tblue{U-HLM}      & \tblue{0.0540} & \tblue{0.0276} & \tblue{0.0559} & \tblue{0.0305} \\
        \tblue{CU-HLM (Offline)} 
                           & \tblue{0.0501} & \tblue{0.0250} & \tblue{0.0530} & \tblue{0.0282} \\
        \tblue{CU-HLM (Online)}  
                           & \tblue{0.0526} & \tblue{0.0273} & \tblue{0.0549} & \tblue{0.0293} \\
        \bottomrule
    \end{tabular}
    }
    \label{tab:additional_metrics_dolly}
\end{table}

\begin{figure}[t]
\centering
\includegraphics[width=\linewidth]{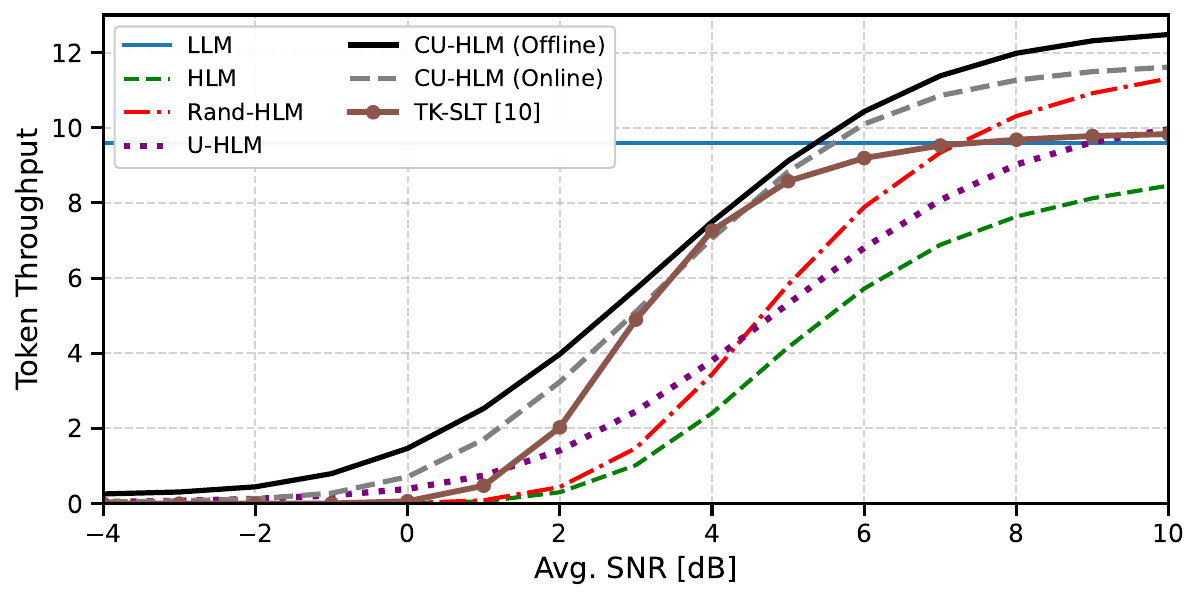}
\caption{\tblue{Token throughput under Rayleigh outage channel.}}
\label{fig:wireless}
\end{figure}

\section{\tblue{Additional Token-Level Similarity Metrics}} \label{append:c}

\tblue{To further validate consistency across evaluation metrics,
we report token-level F1 and Jaccard similarity on the Dolly 15K dataset,
as shown in Table~\ref{tab:additional_metrics_dolly}.}

\tblue{The relative trends are consistent with the inference accuracy results:
Rand-HLM exhibits noticeable degradation,
while U-HLM maintains performance close to HLM.
Both CU-HLM (Offline and Online) achieve competitive similarity scores.}

\section{\tblue{Token Throughput under Outage Channel}} \label{append:d}
\vspace{-3pt}
\tblue{We consider a block fading uplink model with coherence time $\tau = 1$\,ms,
where the channel remains constant within each slot and varies across slots.
An outage-based decoding rule is applied: a transmission is successfully decoded
only if $\text{SNR}(t) \ge \text{SNR}_{\text{target}}$.}

\tblue{Under this model, the accumulated successfully received bits over $T$ slots are
\begin{equation}
B_{\text{RX}}(T)
\!=\! \tau \sum_{t=1}^{T}
\mathbf{1}\big(\text{SNR}(t) \ge \text{SNR}_{\text{target}}\big)
W \log_2\!\left(1\!+\!\text{SNR}_{\text{target}}\right)\!,
\end{equation}
where $W$ is the uplink bandwidth.}

\tblue{The transmission latency for $B$ bits is defined as the minimum $T$
such that $B_{\text{RX}}(T) \ge B$.
If decoding fails in a slot, retransmission occurs in the next slot.
To bound the delay, at most $T_{\max}$ slots are allowed,
and a latency outage is declared if this limit is exceeded.}

\tblue{We conduct experiments with $W = 10$\,MHz and $T_{\max} = 200$\,ms.
Although overall token throughput decreases relative to the idealized channel model,
the relative performance ordering among methods remains consistent.
As shown in Fig.~\ref{fig:wireless}, the proposed designs continue to achieve systematic token throughput gains,
indicating robustness under fading, retransmission, and latency constraints.}

\bibliographystyle{ieeetr}
\bibliography{main}

\end{document}